\documentclass[trackchanges]{aastex63}
\shorttitle{PSP High Beta SSW}
\shortauthors{Huang et al.}
\graphicspath{{./}{figures/}}
\begin{document}

\title{Parker Solar Probe Observations of High Plasma Beta Solar Wind from Streamer Belt}

\correspondingauthor{Jia Huang}
\email{huangjia.sky@gmail.com}

%%%%%%%%%%%%%%%%%%%%%%%%%%%%%%%
\author[0000-0002-9954-4707]{Jia Huang}
\affiliation{Space Sciences Laboratory, University of California, Berkeley, CA 94720, USA.}

\author[0000-0002-7077-930X]{J. C. Kasper}
\affiliation{BWX Technologies, Inc., Washington DC 20001, USA.}
\affiliation{Climate and Space Sciences and Engineering, University of Michigan, Ann Arbor, MI 48109, USA.}

\author[0000-0001-5030-6030]{Davin E. Larson}
\affiliation{Space Sciences Laboratory, University of California, Berkeley, CA 94720, USA.}

\author[0000-0001-6077-4145]{Michael D. McManus}
\affiliation{Space Sciences Laboratory, University of California, Berkeley, CA 94720, USA.}

\author[0000-0002-7287-5098]{P. Whittlesey}
\affiliation{Space Sciences Laboratory, University of California, Berkeley, CA 94720, USA.}

\author[0000-0002-0396-0547]{Roberto Livi}
\affiliation{Space Sciences Laboratory, University of California, Berkeley, CA 94720, USA.}

\author[0000-0002-0396-0547]{Ali Rahmati}
\affiliation{Space Sciences Laboratory, University of California, Berkeley, CA 94720, USA.}

\author[0000-0002-4559-2199]{Orlando Romeo}
\affiliation{Space Sciences Laboratory, University of California, Berkeley, CA 94720, USA.}

\author[0000-0001-6038-1923]{K. G. Klein}
\affiliation{Lunar and Planetary Laboratory, University of Arizona, Tucson, AZ 85719, USA.}

\author[0000-0001-5260-658X]{Weijie Sun}
\affiliation{Climate and Space Sciences and Engineering, University of Michigan, Ann Arbor, MI 48109, USA.}

\author[0000-0001-5260-3944]{Bart van der Holst}
\affiliation{Climate and Space Sciences and Engineering, University of Michigan, Ann Arbor, MI 48109, USA.}

\author[0000-0003-1674-0647]{Zhenguang Huang}
\affiliation{Climate and Space Sciences and Engineering, University of Michigan, Ann Arbor, MI 48109, USA.}

\author[0000-0002-6849-5527]{Lan K. Jian}
\affiliation{Heliophysics Science Division, NASA Goddard Space Flight Center, Greenbelt, MD 20771, USA}

\author[0000-0003-3255-9071]{Adam Szabo}
\affiliation{Heliophysics Science Division, NASA Goddard Space Flight Center, Greenbelt, MD 20771, USA}

\author[0000-0003-1138-652X]{J. L. Verniero}
\affiliation{Heliophysics Science Division, NASA Goddard Space Flight Center, Greenbelt, MD 20771, USA}

\author[0000-0003-4529-3620]{C. H. K. Chen}
\affiliation{School of Physics and Astronomy, Queen Mary University of London, London E1 4NS, UK.}

\author[0000-0001-6807-8494]{B. Lavraud}
\affiliation{Institut de Recherche en Astrophysique et Planétologie, CNRS, UPS, CNES, Université de Toulouse, Toulouse, France}

\author[0000-0003-2981-0544]{Mingzhe Liu}
\affil{LESIA, Observatoire de Paris, Université PSL, CNRS, Sorbonne Université, Université de Paris, 5 place Jules Janssen, 92195 Meudon, France.}

\author[0000-0002-6145-436X]{Samuel T. Badman}
\affil{Smithsonian Astrophysical Observatory, Cambridge, MA 02138 USA.}

\author[0000-0001-6692-9187]{Tatiana Niembro}
\affiliation{Smithsonian Astrophysical Observatory, Cambridge, MA 02138 USA.}

\author[0000-0002-5699-090X]{Kristoff Paulson}
\affiliation{Smithsonian Astrophysical Observatory, Cambridge, MA 02138 USA.}

\author[0000-0002-7728-0085]{M. Stevens}
\affiliation{Smithsonian Astrophysical Observatory, Cambridge, MA 02138 USA.}

\author[0000-0002-3520-4041]{A. W. Case}
\affiliation{Smithsonian Astrophysical Observatory, Cambridge, MA 02138 USA.}

\author[0000-0002-1573-7457]{Marc Pulupa}
\affiliation{Space Sciences Laboratory, University of California, Berkeley, CA 94720, USA.}

\author[0000-0002-1989-3596]{Stuart D. Bale}
\affil{Physics Department, University of California, Berkeley, CA 94720-7300, USA.}
\affil{Space Sciences Laboratory, University of California, Berkeley, CA 94720, USA.}
\affil{The Blackett Laboratory, Imperial College London, London, SW7 2AZ, UK.}
\affil{School of Physics and Astronomy, Queen Mary University of London, London E1 4NS, UK.}

\author[0000-0001-5258-6128]{J. S. Halekas}
\affil{Department of Physics and Astronomy, University of Iowa, Iowa City, IA 52242, USA.}

%%%%%%%%%%%%%%%%%%%%%%%%%%%%%%%%%%%%%%%%%%%%%%%%%%%%%%%%%
\begin{abstract}
In general, slow solar wind from the streamer belt forms a high plasma beta equatorial plasma sheet around the heliospheric current sheet (HCS) crossing, namely the heliospheric plasma sheet (HPS). Current Parker Solar Probe (PSP) observations show that the HCS crossings near the Sun could be full or partial current sheet crossing (PCS), and they share some common features but also have different properties. In this work, using the PSP observations from encounters 4 to 10, we identify streamer belt solar wind from enhancements in plasma beta, and we further use electron pitch angle distributions to separate it into HPS solar wind that around the full HCS crossings and PCS solar wind that in the vicinity of PCS crossings. Based on our analysis, we find that the PCS solar wind has different characteristics as compared with HPS solar wind: a) PCS solar wind could be non-pressure-balanced structures rather than magnetic holes, and the total pressure enhancement mainly results from the less reduced magnetic pressure; b) some of the PCS solar wind are mirror unstable; c) PCS solar wind is dominated by very low helium abundance but varied alpha-proton differential speed. We suggest the PCS solar wind could originate from coronal loops deep inside the streamer belt, and it is pristine solar wind that still actively interacts with ambient solar wind, thus it is valuable for further investigations on the heating and acceleration of slow solar wind.  
\end{abstract}


\section{Introduction} \label{sec:intro}

Parker Solar Probe (PSP) aims to enter the atmosphere of the Sun and provides in-situ measurements to uncover the properties of solar wind close to its source regions \citep{Fox-2016}. The PSP has completed its initial fourteen orbits by December 2022, with the deepest perihelion reaching a heliocentric distance of about 13.3 solar radii ($R_S$), and it entered the solar corona for the first time on 28 April 2021 \citep{kasper-2021}. PSP has many extraordinary observations, and the new data give us a chance to investigate the properties of streamer belt solar wind near the Sun.   

The origin and evolution of slow solar wind are still debatable, and the multiple source regions of the slow solar wind are one of the difficulties. The streamer belt is believed to be a certain source of the slow solar wind, thus it is suitable to study the nature of slow solar wind from a specific source region with less uncertainty. Solar wind from streamer belt generally forms a low speed but high plasma beta ($\beta$) solar stream region, i.e. heliospheric plasma sheet (HPS), which always embeds a heliospheric current sheet (HCS) \citep[e.g.][]{borrini-1981, winterhalter-1994, smith-2001, crooker-2004HPS, suess-2009, Liu-2014, huang-2016a}. However, current PSP observations reveal that HCS crossings and HPS solar winds in the near Sun environment are much more dynamic than that at 1 AU, for example the HCSs have multiple crossings \citep{szabo-2019, lavraud-2020, phan-2020}, the magnetic reconnections are prevalent around the HCS crossings \citep{lavraud-2020, phan-2020, phan-2021}, multiple small scale structures are found in HPS solar winds \citep{szabo-2019, lavraud-2020, rouillard-2020, zhao-2021, reville-2022}. 

Moreover, \citet{lavraud-2020} and \citet{phan-2021} find that there is a category of partial current sheet crossings (PCSs), which are different from the well known HCS crossings that defined as fully crossings of two magnetic field sectors with different polarities. The PCSs stay at the same magnetic field sector without crossing the sector boundary. They generally appear in the vicinity of HCS crossings and also show signature of density enhancement, but they exhibit more significant changes in suprathermal electrons and have smaller magnetic field rotation as compared with HCS crossings, and sometimes they display reconnection jet signatures \citep{lavraud-2020, phan-2021}. The PCSs could be caused by the warped HCSs \citep[e.g.][]{peng-2019}, but their signatures of long duration and recurrent appearance imply that they are most likely generated by the traveling large plasma blobs bulging onto both sides of the HCS crossings \citep[e.g.][]{phan-2004, lavraud-2020, phan-2021}. As a result, the PCSs represent the pristine state of the solar wind from the streamer belt, and it is valuable to compare the PCS solar wind with the HCS solar wind (i.e. HPS) to infer their differences on kinetic properties and also origins.  

In this work, we identify HPS solar wind and PCS solar wind from encounter 4 to encounter 10 (E4-E10), and we then compare their pressures, temperature anisotropies, and helium signatures to infer their different behaviors and origins. In section \ref{sec:data}, we introduce the data we used in this work. Section \ref{sec:results} presents the results in E4, whereas section \ref{sec:Eallresults} shows the multi-event analysis from E4 to E10. The discussion and summary are included in section \ref{sec:disc} and section \ref{sec:sum}, respectively.

\section{Data} \label{sec:data}
The instrument suites of Solar Wind Electrons, Alphas, and Protons (SWEAP)  \citep{kasper-2016} and FIELDS \citep{bale-2016} onboard PSP provide the data used in this work. The SWEAP includes the Solar Probe Cup (SPC) \citep{Case-2020}, Solar Probe Analyzer for Electrons (SPAN-E) \citep{whittlesey-2020}, and Solar Probe Analyzer for Ions (SPAN-I) \citep{livi-2022}. The SWEAP is designed to measure the velocity distributions of solar wind electrons, alpha particles, and protons \citep{kasper-2016}. FIELDS is designed to measure DC and fluctuation magnetic and electric fields, plasma wave spectra and polarization properties, the spacecraft floating potential, and solar radio emissions \citep{bale-2016}.

In this work, we use the electron data from SPAN-E, and the magnetic field data from the FIELDS. The electron density is derived from the analysis of plasma quasi-thermal noise (QTN) spectrum measured by the FIELDS Radio Frequency Spectrometer \citep{pulupa-2017, moncuquet-2020}. 
The fitted proton and alpha data from SPAN-I are used to study the alpha associated signatures. We also select the best SPAN-I data and SPC data to calculate the radial power law indices of pressures as shown in the appendix. The temperature components are retrieved from bi-Maxwellian fitting to the proton channel spectra observed by SPAN-I.
SPAN-I measures three-dimensional (3D) velocity distribution functions of the ambient ions in the energy range from several eV/q to 20 keV/q with a maximum time resolution of 0.437 s, and it has a time of flight section that enables it to differentiate the ion species \citep{kasper-2016}. The details of the fitted proton and alpha data are described in \citet{livi-2022}, \citet{finley-2020} and \citet{mcmanus-2022}. But the SPAN-I measurements used here are from low cadence downlinked data, and the cadence of the fitted proton and alpha data are 6.99 s and 13.98 s, respectively \citep{finley-2020, verniero-2020, mcmanus-2022}. The FIELDS instrument collects high resolution vector magnetic fields with variable time resolutions. The 4 samples per cycle (i.e. 4 samples per 0.874 s) data are used here. 

\section{E4 Results \label{sec:results}}

\subsection{Overview of high plasma beta solar wind in E4 \label{sec:overview}}
\begin{figure}
\epsscale{0.95}
\plotone{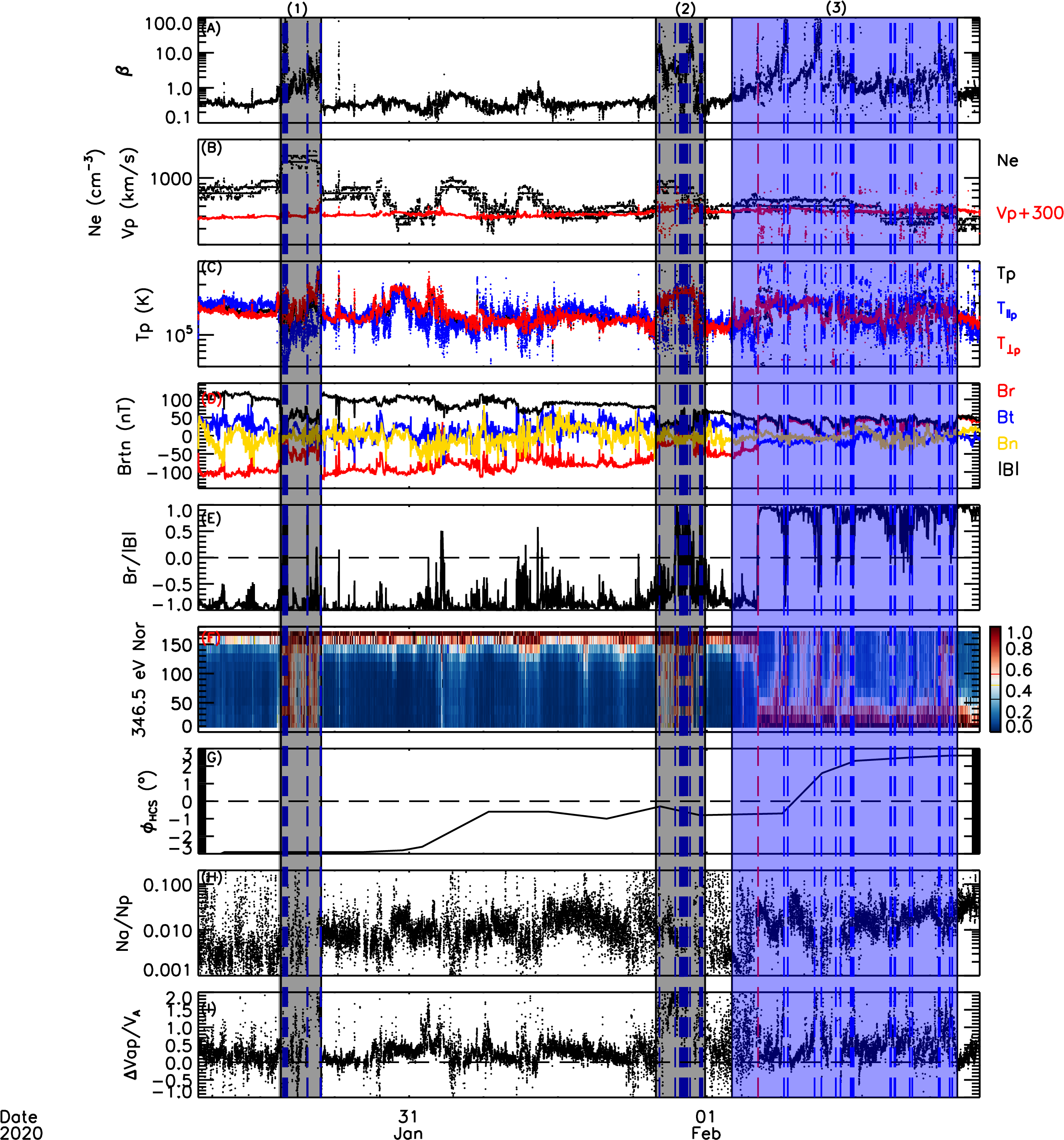}
\caption{Overview of the high plasma beta streamer belt solar wind in E4. From top to bottom, the panels show plasma beta $\mathrm{\beta}$, electron number density $\mathrm{Ne}$ and solar wind speed $\mathrm{Vp}$, proton temperatures, magnetic field components in RTN coordinates, radial magnetic field to total magnetic field strength ratio $\mathrm{Br/|B|}$, normalized pitch angle distribution of suprathermal electrons (e-PAD) at energy of 346.5 eV, angular distance to the HCS crossing $\mathrm{\phi_{HCS} (^o)}$, alpha to proton abundance ratio $\mathrm{Na/Np}$, and alpha-proton differential speed normalized by local Alfvén speed $\mathrm{\Delta Vap/V_A}$. In panel (B), we add 300 km/s to solar wind speed. In panel (C), the perpendicular temperature ($\mathrm{T_{\perp p}}$), parallel temperature ($\mathrm{T_{\parallel p}}$), and total temperature ($\mathrm{T_p}$) are indicated by red, blue, and black dots, respectively. The gray shaded regions (1) and (2) mark high plasma beta solar winds around partial current sheet crossings. The blue shaded region (3) is HPS solar wind around the HCS crossing as indicated by the red vertical dashed line. The blue vertical dashed lines represent the switchbacks in these high beta solar winds. 
} \label{fig:overview}
\end{figure}
%%%%%%%%%%%%%%%%%%%%

Figure \ref{fig:overview} presents an overview of high plasma beta solar wind in E4 between 2020-01-30 12:00 UT and 2020-02-02 00:00 UT. From top to bottom, the panels show plasma beta $\mathrm{\beta}$, QTN electron number density $\mathrm{Ne}$ and solar wind speed $\mathrm{Vp}$, proton temperatures (total temperature $\mathrm{Tp}$, parallel temperature $\mathrm{T_{\parallel p}}$ and perpendicular temperature $\mathrm{T_{\perp p}}$), magnetic field components in RTN coordinates, radial magnetic field to total magnetic field strength ratio $\mathrm{Br/|B|}$, normalized pitch angle distribution of suprathermal electrons (e-PAD) at energy of 346.5 eV, angular distance to the HCS crossing $\mathrm{\phi_{HCS}\ (^o)}$, alpha to proton abundance ratio $\mathrm{Na/Np}$, and alpha-proton differential speed normalized by local Alfvén speed $\mathrm{\Delta Vap/V_A}$. 
The three shaded regions mark the high $\mathrm{\beta}$ solar winds, with the blue dashed lines indicating the middle times of switchbacks identified with our algorithm \citep{kasper-2019, huang-2023SB}. $\mathrm{\beta = 2\mu_0 n_p k_B T_p/B^2}$, where $\mathrm{\mu_0}$ is vacuum magnetic permeability, $\mathrm{k_B}$ is Boltzmann constant, $\mathrm{B}$ is the magnetic field strength, $\mathrm{n_p}$ and $\mathrm{T_p}$ are the number density and temperature of protons, respectively. 
The red dashed vertical line in region (3) suggests the HCS crossing, when the $\mathrm{Br/|B|}$ changes polarity (panel (E)) and the pitch angle distribution of suprathermal electrons changes direction (panel (F)) simultaneously. 

In this work, we use high $\beta$, which is larger than 1 and also larger than that in the ambient solar wind, as the primary signature to identify the streamer belt solar wind. We keep using HPS solar wind to name the solar wind in the vicinity of full HCS crossings, whereas we define the solar wind around PCSs as the PCS solar wind. As stated in the introduction, we further combine the magnetic field polarity and e-PAD to indicate full HCS or PCS crossings, i.e. the magnetic field polarity and e-PAD both change directions before and after full HCS crossings, whereas the magnetic field polarity doesn't change but e-PAD significantly scattered around PCS crossings. Generally, we search for PCS crossings in about two days before and after HCS crossings.  
Therefore, regions (1) and (2) are PCS solar wind, and region (3) is HPS solar wind, which is consistent with the classification in \citet{phan-2021}. The solar wind speed is less than 300 km/s during this time period and it doesn't change a lot inside and outside the three regions. The e-PAD in panel (F) changes direction in HPS solar wind, but it stays predominantly in the same direction and scatters a lot in PCS solar wind, inferring complicated physical processes may involve with \citep{halekas-2021}. We note the HPS solar wind is slightly closer to the HCS crossing than the PCS solar wind, as indicated by the angular distance to the HCS crossing $\mathrm{\phi_{HCS}}$ that derived from the Potential-field Source-surface (PFSS) model \citep{szabo-2019, badman-2020, stansby-2020, chen-2021}. In these regions, we can see the $\mathrm{Ne}$ and temperatures increase significantly but the magnetic field strength $\mathrm{|B|}$ decreases greatly at the same time, thus the $\beta$ enhances profoundly as compared with ambient solar wind, and sometimes it could even reach to about 1000. Some of the super high $\beta$ solar wind involves with the switchbacks as shown by the blue dashed lines, due to the magnetic field reversals lead to very small magnetic pressures. However, there is still some super high $\beta$ solar wind shows no relation with switchbacks, and this kind of solar wind seems to have low $\mathrm{Na/Np}$, implying the solar streams should come from the coronal loops deep inside the streamer belt \citep{suess-2009}. Moreover, in the HPS and PCS solar winds, $\mathrm{T_{\parallel p}}$ and $\mathrm{T_{\perp p}}$ also increase, implying their thermal states may not be stable, and the $\mathrm{\Delta Vap/V_A}$ deviates from zero, indicating the solar wind may still under evolution.

\subsection{Pressure variations \label{sec:PBS}}

\begin{figure}
\epsscale{1.0}
\plotone{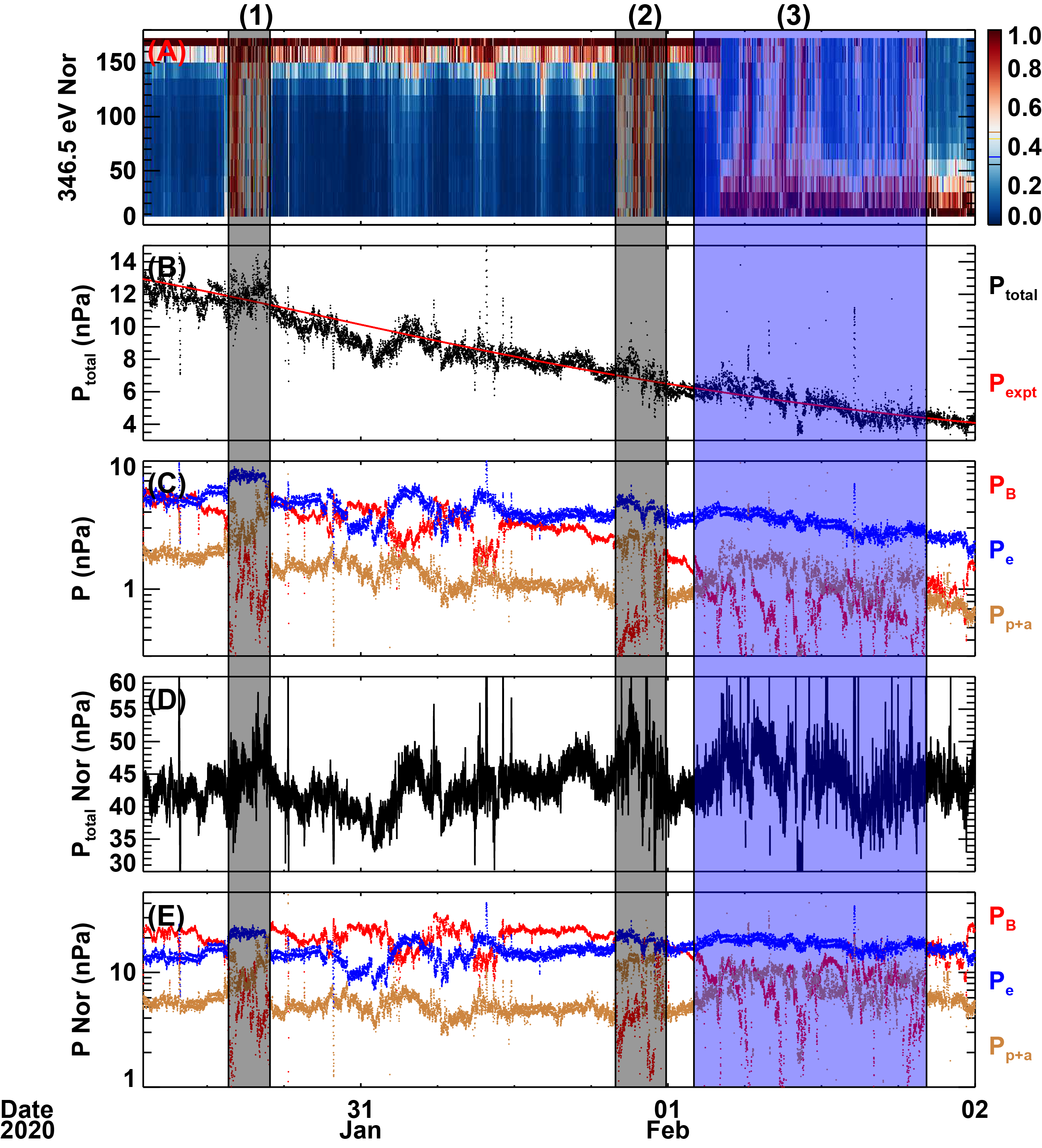}
\caption{The pressure variations in E4 streamer belt solar wind. From top to bottom, the panels show the normalized e-PAD at energy of 346.5 eV, total pressure ($\mathrm{P_{total}}$) and expected total pressure ($\mathrm{P_{expt}}$), pressure components ($\mathrm{P_{B}}$, $\mathrm{P_{e}}$, $\mathrm{P_{p+a}}$), normalized total pressure, and normalized pressure components. The three shaded regions and the blue dashed lines mark the high beta streamer belt solar wind and switchbacks, respectively, as that in Figure \ref{fig:overview}.
}\label{fig:pressure}
\end{figure}
%%%%%%%%%%%%%%%%%%%%%%

Thermal pressure gradients drive the solar wind flow out from the solar corona \citep{cranmer-2019, owens-2020}. In general, the pressure-balanced structures are normal in the interplanetary space, such as tangential discontinuities, rotational discontinuities, magnetic holes, small transients, magnetic reconnection exhausts, and so on \citep{belcher-1969, belcher-1971, burlaga-1971, burlaga-1990b, wei-2006, stevens-2007, yu-2014, mistry-2017}. Moreover, the total pressure of HPS is comparable to that in the ambient solar wind, and the small scale structures inside HPSs also show a pressure-balanced signature \citep{burlaga-1990b, winterhalter-1994, crooker-2004HPS, yamada-2010, foullon-2011, yu-2014}. However, the non-pressure-balanced structures are rare in the interplanetary medium except for large magnetic clouds that in expansions and the co-rotating interaction regions that are formed by compressions. In addition, interplanetary shock fronts and magnetic cloud boundary layers are also found to be non-pressure-balanced \citep{wei-2006, zuo-2006, wangyi-2010, priest-2014, zhou-2018, zhou-2019}. The interplanetary shock front is a relatively thin transitional layer from the quasi-uniform solar wind to the disturbed solar wind, and the interactions within the shock front are efficient to convert the flow energy into thermal energy and accelerate particles to significant energies \citep[e.g.,][]{priest-2014, sapunova-2017}. The boundary layers of magnetic clouds are formed as the magnetic clouds interact with ambient solar wind during propagation, and they have complicated fine structures like slow shock, magnetic reconnection exhaust, magnetic field reversal, and enhanced wave activity, implying the boundary layer is sufficient to heat and accelerate the encountered solar wind \citep{wei-2003, wei-2006, zuo-2006, wangyi-2010, priest-2014, zhou-2018, zhou-2019}. 
Therefore, pressure is an important indicator of solar wind states, and non-pressure-balanced signature always associates with crucial physical processes like plasma heating and acceleration, plasma wave activity, and magnetic reconnections. As PSP dives into the solar atmosphere, it has chances to observe more pristine solar winds that still actively interact with ambient solar wind. Consequently, it is essential to investigate the pressure variations in the streamer belt solar wind in the inner heliosphere, which may shed light on the long-standing mysteries of slow solar wind in terms of the formation, evolution, heating and acceleration processes.

Figure \ref{fig:pressure} shows the pressure variations in E4 streamer belt solar wind. From top to bottom, the panels show the normalized e-PAD at energy of 346.5 eV, total pressure ($\mathrm{P_{total}}$) and expected total pressure ($\mathrm{P_{expt}}$), pressure components ($\mathrm{P_{B}}$, $\mathrm{P_{e}}$, $\mathrm{P_{p+a}}$), normalized total pressure ($\mathrm{P_{total}\ Nor}$), and normalized pressure components. The three shaded regions mark the high $\mathrm{\beta}$ streamer belt solar wind as that in Figure \ref{fig:overview}. The definitions of magnetic pressure $\mathrm{P_{B}}$, electron pressure $\mathrm{P_{e}}$, proton and alpha pressure $\mathrm{P_{p+a}}$, and total pressure $\mathrm{P_{total}}$ are presented in Appendix \ref{sec:parameters}. Since the pressures vary with heliocentric distances and PSP has an elliptic orbit, we investigate their radial evolutions in Appendix \ref{sec:presevo}. We further use the derived functions to normalize the pressures to 20 $R_S$ as shown in panels (D) and (E), and also to estimate the expected total pressure as a function of radial distance $\mathrm{P_{expt} = 10^{5.96} \times R^{-3.31}\ (nPa)}$ in panel (B), where $\mathrm{R}$ is the heliocentric distance in unit of $\mathrm{R_S}$. From panel (B), we can see $\mathrm{P_{total}}$ changes about threefold in about two days, indicating it is necessary to normalize the pressures when studying pressure associated signatures. 

In the high $\mathrm{\beta}$ streamer belt solar wind, we can see the total pressure enhances as compared with ambient solar wind. The enhancements could be seen both from the $\mathrm{P_{total}\ Nor}$ in panel (D) and from the comparison between $\mathrm{P_{total}}$ and $\mathrm{P_{expt}}$ in panel (B). In the PCS solar wind, the $\mathrm{P_{total}\ Nor}$ increases about 15\% and 25\% in regions (1) and (2), respectively. In panel (E), we can see the magnetic pressure decreases significantly in the PCS solar wind, whereas the thermal pressure increases more profoundly, resulting in the enhancement of total pressure. In addition, we note the total pressure also shows signature of enhancement in the HPS solar wind, with the variations of pressure components similar to that in the PCS solar wind. This implies that the HPS solar wind in the inner heliosphere may be more active than that observed at 1 AU and beyond. However, our results in the following section \ref{sec:EallPressure} indicate that the pressure enhancements are much more distinctive in PCS solar wind than that in HPS solar wind, indicating the PCS solar wind could be non-pressure-balanced structures. This characteristic suggests the PCS solar wind in the inner heliosphere could be pristine solar wind that still actively interacts with ambient solar wind. 

Besides, we can see the switchbacks, as marked by blue dashed lines, could modify the total pressure, i.e. the crash of magnetic pressure leads to the temporary decrease of total pressure inside the streamer belt solar wind. But the total pressure of switchbacks seem to be comparable to the ambient solar wind outside the high plasma beta solar wind, implying the switchbacks are roughly pressure-balanced structures as suggested by \citet{bale-2021}.

\subsection{Temperature anisotropy characteristics \label{sec:Tanis}}

\begin{figure}
\epsscale{1.0}
\plotone{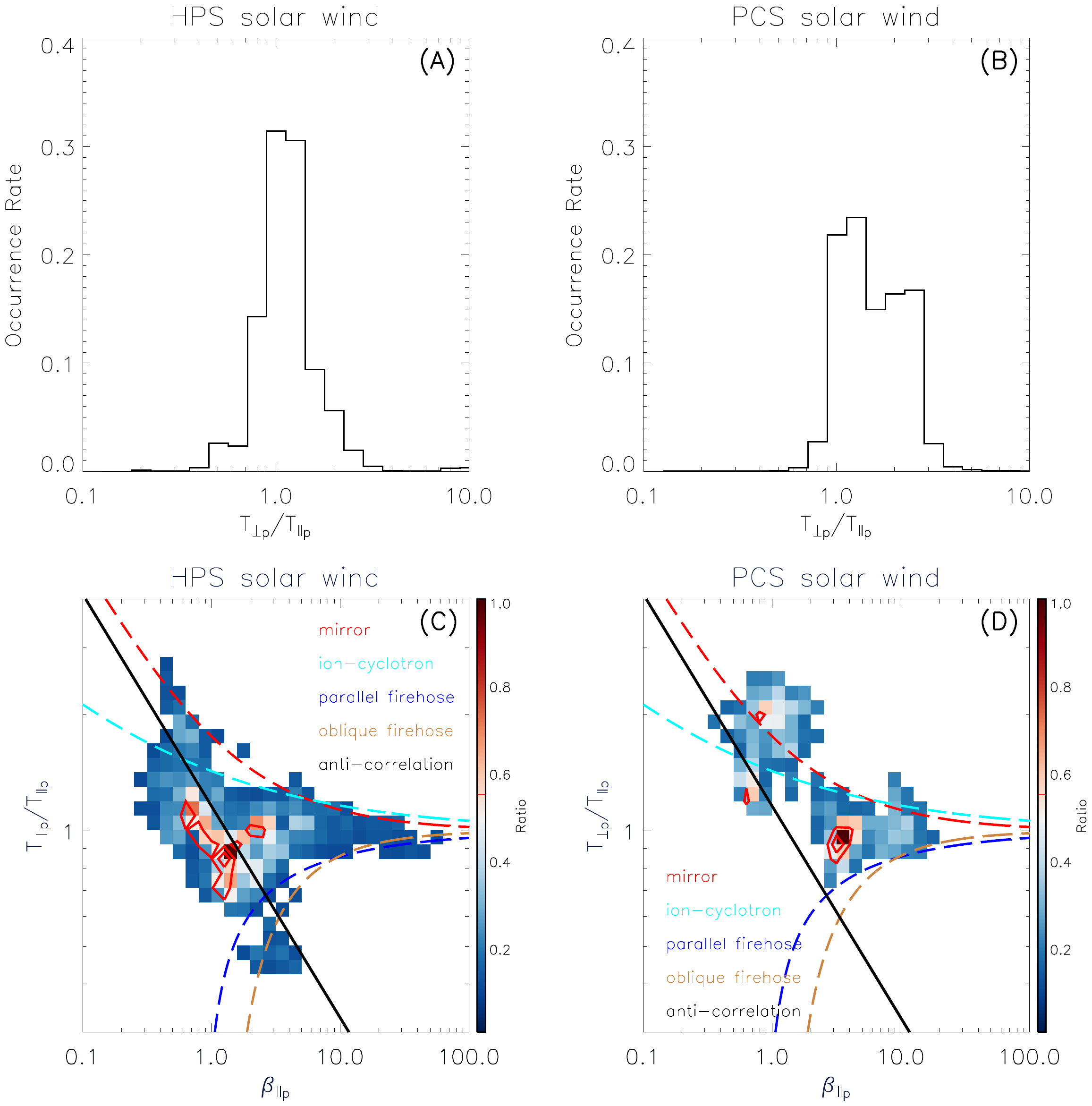}
\caption{Temperature anisotropies in E4 streamer belt solar wind. Panels (A) and (C) display the temperature anisotropy ($\mathrm{T_{\perp p}/T_{\parallel p}}$) signatures in HPS solar wind, and panels (B) and (D) show same parameters in PCS solar wind. Panels (A) and (B) present the occurrence rates of temperature anisotropies in HPS and PCS solar winds, respectively. Panels (C) and (D) exhibit the temperature anisotropy versus the parallel plasma beta ($\mathrm{\beta_{\parallel p}}$) in HPS and PCS solar winds, respectively. The different color dashed lines are instabilities as indicated by the legend, and the black line is anti-correlation relationship. The red contours indicate 50\% and 75\% measurements. 
} \label{fig:Tani}
\end{figure}
%%%%%%%%%%%%%%

The thermodynamic property is pivotal to understanding the kinetic processes governing the dynamics of interplanetary medium \citep{kasper-2002, kasper-2003, he-2013, maruca-2013, huang-2020}. Temperature anisotropy ($\mathrm{T_{\perp p}/T_{\parallel p}}$) arises when $\mathrm{T_{\perp p}}$ and/or $\mathrm{T_{\parallel p}}$ departs from thermal equilibrium, which indicates anisotropic heating and/or cooling processes act preferentially in one direction \citep{maruca-2011}, and such preferential heating/cooling is supported by the observed departures of $\mathrm{T_{\perp p}/T_{\parallel p}}$ from adiabatic predictions in solar wind observations \citep{gary-1993, matteini-2007}. As temperature anisotropy departs from unity, anisotropy-driven instabilities such as mirror, ion-cyclotron, parallel and oblique firehose instabilities arise, and act to isotropize the plasma \citep{gary-1993, maruca-2011}. Therefore, a study of the temperature anisotropy characteristics in PCS solar wind and HPS solar wind is helpful to differentiate the dynamic processes therein.

Figure \ref{fig:Tani} shows the characteristics of temperature anisotropies in E4 streamer belt solar wind. The left panels (A) and (C) display the $\mathrm{T_{\perp p}/T_{\parallel p}}$ signatures in HPS solar wind, whereas the right two panels show the same parameters in PCS solar wind. Panels (A) and (B) present the occurrence rates of temperature anisotropies in HPS and PCS solar winds, respectively. Panels (C) and (D) exhibit the $\mathrm{T_{\perp p}/T_{\parallel p}}$ versus the parallel plasma beta ($\mathrm{\beta_{\parallel p}}$) in HPS and PCS solar winds, respectively. The colored dashed lines are instabilities as indicated by the legend with the thresholds from \citet{hellinger-2006}, and the black line is anti-correlation relationship between $\mathrm{T_{\perp p}/T_{\parallel p}}$ and $\mathrm{\beta_{\parallel p}}$ derived by \citet{marsch-2004}. 

In the HPS solar wind, we can see the $\mathrm{T_{\perp p}/T_{\parallel p}}$ is almost isotropic ($\sim$ 1) from panel (A), and the $\mathrm{T_{\perp p}/T_{\parallel p}}$ \textit{versus} $\mathrm{\beta_{\parallel p}}$ distribution is well limited by the instabilities as indicated in panel (C). These signatures suggest the HPS solar wind is generally in thermal equilibrium state. However, the PCS solar wind shows different signatures. On one hand, PCS solar wind seems to have two populations as shown by the right panels. One population has isotropic temperature anisotropy, whereas the other population is much more anisotropic. On the other hand, the isotropic population is well limited by the instabilities, but part of the anisotropic population is beyond the mirror instability constraint. This result implies that some of the PCS solar wind is further heated, especially in perpendicular direction to the background magnetic field if we combine the temperature observations in Figure \ref{fig:overview}(C). The possible heating mechanisms could be the prevalent magnetic reconnections around current sheet crossings \citep{lavraud-2020, phan-2020, phan-2021}, the turbulence in inner heliosphere \citep{chen-2021}, the switchbacks \citep{akhavan-2022}, and/or other processes. Among these mechanisms, magnetic reconnection is the most likely heating mechanism due to its prevalence and efficiency to heat the plasma. As indicated by \citet{chen-2021}, the E4 streamer belt solar wind shows much lower Alfvénic turbulence energy flux, which may not be able to accelerate and heat the solar wind. In addition, the switchbacks are both found in HPS solar wind and PCS solar wind, but the isotropic temperatures in HPS solar wind imply that the switchbacks can not heat the solar wind efficiently.

\subsection{Helium signatures \label{sec:Helium}}

\begin{figure}
\epsscale{1.2}
\plotone{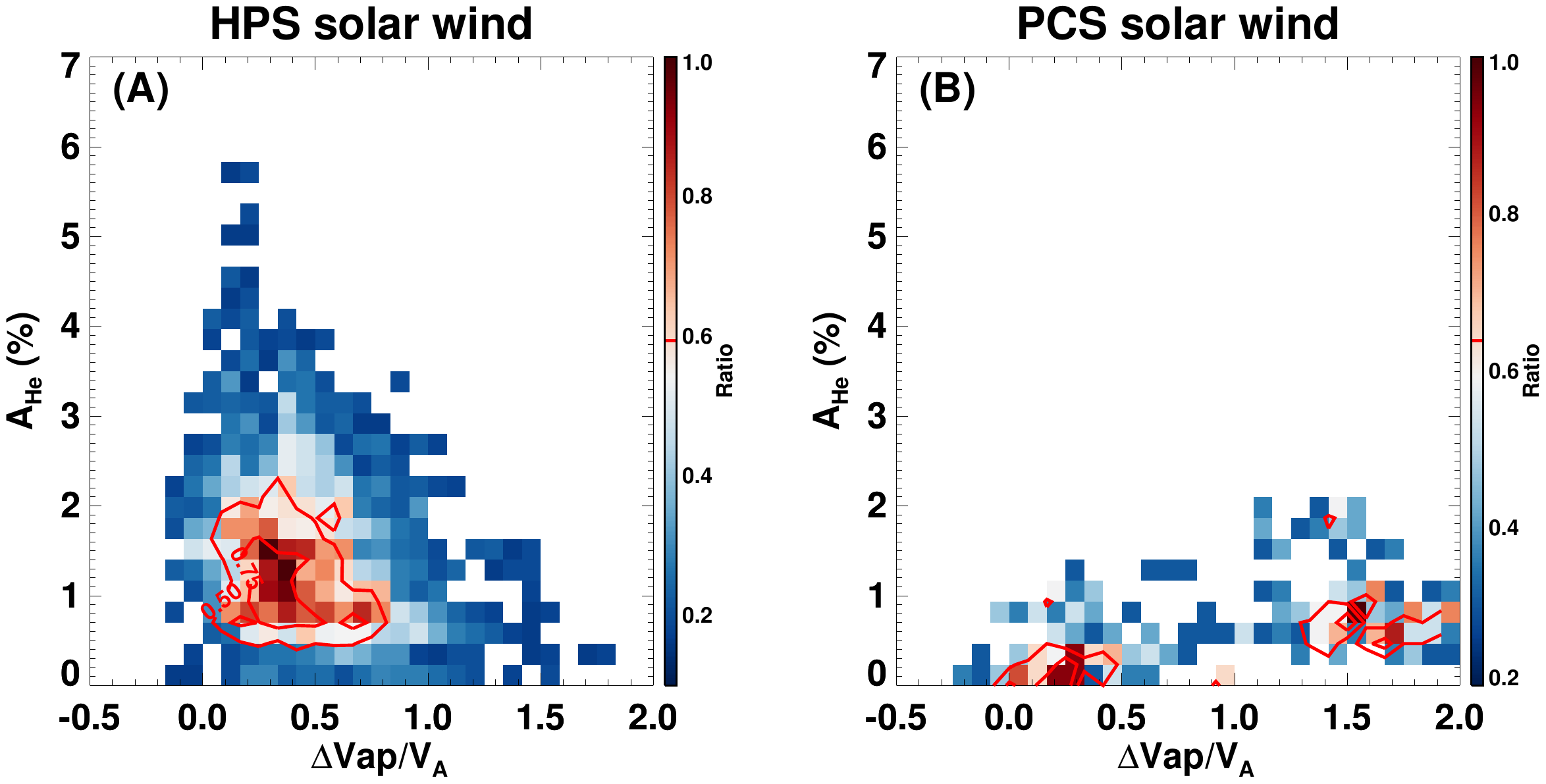}
\caption{The distributions of helium abundance ($\mathrm{A_{He}}$) versus alpha-proton differential speed that normalized by local Alfvén speed ($\mathrm{\Delta V_{\alpha p}/V_A}$) in E4 streamer belt solar wind. Panels (A) and (B) present the helium characteristics in HPS and PCS solar winds, respectively. The colors indicate the occurrence ratios. The red contours in both panels represent 50\% and 75\% of measurements.
} \label{fig:NVap}
\end{figure} 

\begin{figure}
\epsscale{0.8}
\plotone{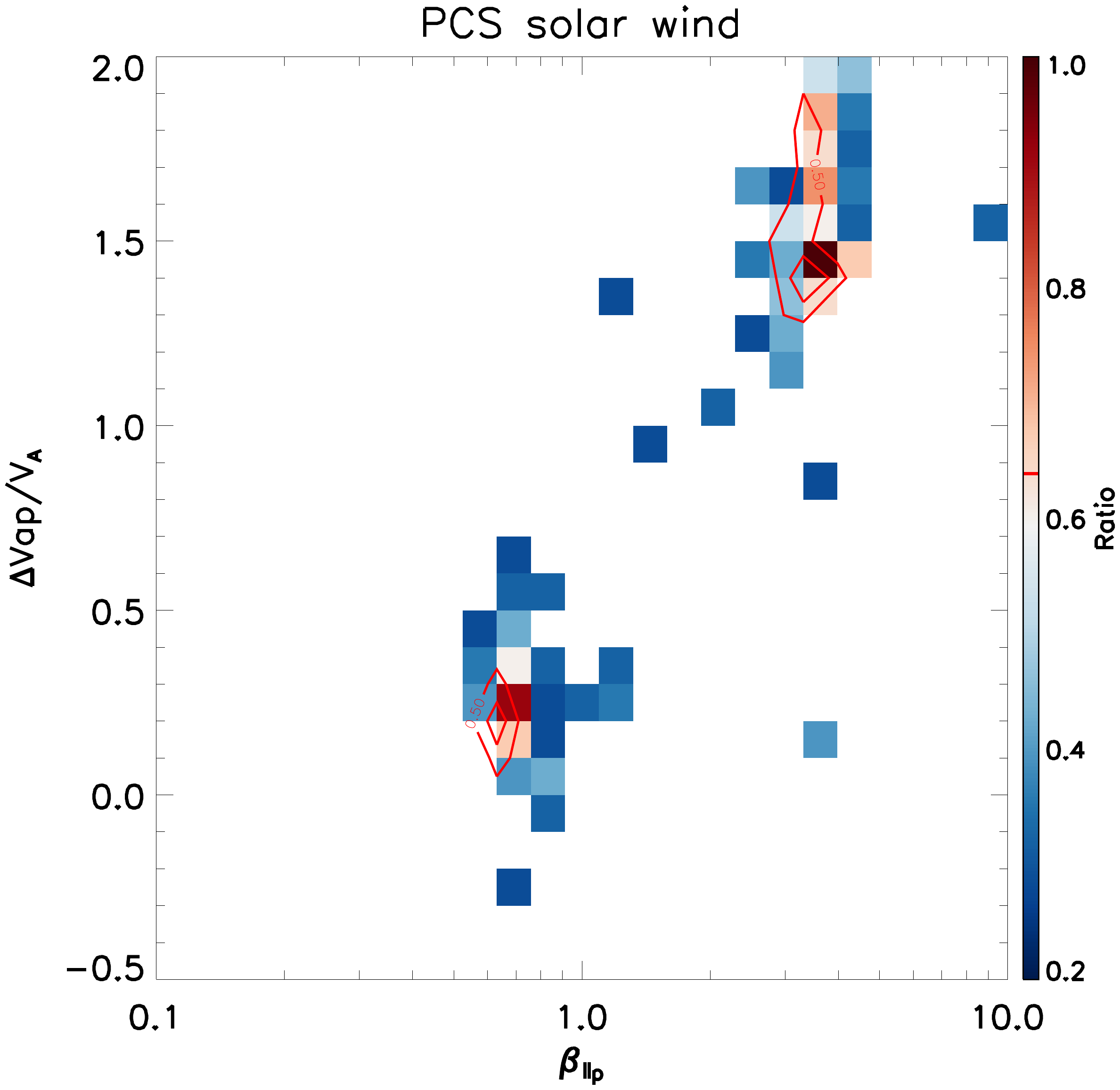}
\caption{The distributions of parallel plasma beta ($\mathrm{\beta_{\parallel p}}$) versus alpha-proton differential speed that normalized by local Alfvén speed ($\mathrm{\Delta V_{\alpha p}/V_A}$) in E4 PCS solar wind. The colors indicate the occurrence ratios. The red contours represent 50\% and 75\% of measurements. 
} \label{fig:betaVap}
\end{figure}
%%%%%%%%%%%%%%%%%%%

The helium signatures connect the in situ solar wind with its source regions at the Sun \citep[e.g.][]{bochsler-2007, aellig-2001, kasper-2012, huang-2016b, huang-2018, fu-2018}. 
In fast solar wind, the helium abundance ratio ($A_{He} = N_\alpha/N_p \times 100\%$) usually increases, and the alpha-proton drift speed ($\mathrm{\Delta V_{\alpha p}}$) is generally large and comparable to local Alfvén speed ($\mathrm{V_A}$), implying the helium-rich population is from open magnetic field regions in the Sun \citep{borrini-1981, gosling-1981, marsch-1982, steinberg-1996, reisenfeld-2001, suess-2009, berger-2011}. However, in slow solar wind, $\mathrm{A_{He}}$ varies with solar activity, i.e. helium-poor population is usually observed at solar minimum, which could originate from streamer belt, but helium-rich population observed at solar maximum is majorly from active regions \citep{kasper-2007, kasper-2012, alterman-2018, alterman-2019}. Additionally, $\mathrm{\Delta V_{\alpha p}}$ is close to zero in slow solar wind \citep{marsch-1982, steinberg-1996, reisenfeld-2001, berger-2011}. Further, studies reveal approximate bimodal distributions of $\mathrm{A_{He}}$ versus $\mathrm{\Delta V_{\alpha p}/V_A}$ in the solar wind observed around 1 AU, with the high $\mathrm{A_{He}}$ and high $\mathrm{\Delta V_{\alpha p}/V_A}$ population probably escaping directly along open magnetic field lines as described by wave-turbulence driven models, whereas low $\mathrm{A_{He}}$ and low $\mathrm{\Delta V_{\alpha p}/V_A}$ population releasing through magnetic reconnection processes \citep[][and references therein]{durovcova-2017, durovcova-2019, fu-2018}. Therefore, the helium signatures could help identify the possible origins of HPS solar wind and PCS solar wind. 

As described in Appendix \ref{sec:parameters}, we follow \citet{reisenfeld-2001} and \citet{fu-2018} to define the $\mathrm{\Delta V_{\alpha p}}$ as the field-aligned differential speed, i.e. $\mathrm{\Delta V_{\alpha p} = (v_{\alpha r}-v_{pr})/cos(\theta)}$, where $\mathrm{v_{\alpha r}}$ and $\mathrm{v_{pr}}$ are the radial speeds of alpha particle and proton, respectively, and $\theta$ measures the angle of the magnetic field vector from radial direction. Here, we further require $\mathrm{cos(\theta) = |B_r/B|}$ to remove its dependency on magnetic field polarity, where $B_r$ represents the radial magnetic field. %In order to reduce measurement uncertainties, we also follow them to require $\mathrm{\theta}$ to be smaller than 72.5 degree.  

Figure \ref{fig:NVap} shows the distributions of $\mathrm{A_{He}}$ versus $\mathrm{\Delta V_{\alpha p}/V_A}$ in HPS solar wind (panel (A)) and PCS solar wind (panel (B)) during E4. The colors indicate the occurrence ratios, whereas the red contours represent 50\% and 75\% of measurements. In HPS solar wind, we can see the plasma is dominated by low $\mathrm{\Delta V_{\alpha p}/V_A}$, but $\mathrm{A_{He}}$ varies from low to high values. The distribution is centered around $\mathrm{A_{He}} = 1.5\%$ and $\mathrm{\Delta V_{\alpha p}/V_A} = 0.35$, implying the HPS solar wind mainly originates from closed magnetic field region via magnetic reconnection processes. But the large $\mathrm{A_{He}}$ population indicates that some of the HPS solar wind comes from the open field region, which may be the leg/flank region of helmet streamer \citep{suess-2009}. In contrast, the PCS solar wind is dominated by low $\mathrm{A_{He}}$ with the $\mathrm{\Delta V_{\alpha p}/V_A}$ changes from small to large values. From panel (B), we can see that there are two populations in PCS solar wind. One population has very low $\mathrm{A_{He}}$ ($\sim 0.2\%$) and small $\mathrm{\Delta V_{\alpha p}/V_A}$ ($\sim 0.25$), suggesting the solar wind is from closed magnetic field region via probably magnetic reconnection. The other population maintains low $\mathrm{A_{He}}$ ($\sim 0.7\%$) but large $\mathrm{\Delta V_{\alpha p}/V_A}$ ($\sim 1.6$), implying the solar wind is also from closed field region but the alpha particles could be preferentially accelerated \citep{isenberg-1983, kasper-2017}. 

In addition, we note the PCS solar wind shows two populations in both Figure \ref{fig:Tani} and Figure \ref{fig:NVap}, thus we would like to know if the two populations are intrinsically related to each other. In Figure \ref{fig:betaVap}, we plot the distribution of $\mathrm{\beta_{\parallel p}}$ versus $\mathrm{\Delta V_{\alpha p}/V_A}$ in E4 PCS solar wind, due to the two parameters could significantly separate the two populations in their distributions. From this figure, we can see $\mathrm{\Delta V_{\alpha p}/V_A}$ is almost linearly associated with $\mathrm{\beta_{\parallel p}}$. Therefore, the low $\mathrm{A_{He}}$ and low $\mathrm{\Delta V_{\alpha p}/V_A}$ population in Figure \ref{fig:NVap} should have low $\mathrm{\beta_{\parallel p}}$ but high $\mathrm{T_{\perp p}/T_{\parallel p}}$ (i.e. the anisotropic population), whereas the low $\mathrm{A_{He}}$ and high $\mathrm{\Delta V_{\alpha p}/V_A}$ population corresponds to the isotropic population in Figure \ref{fig:Tani}. This is a really interesting result. 

In combination of all the observations, we can infer that the non-pressure-balanced PCS solar wind has two populations. One population shows very low $\mathrm{A_{He}}$, low $\mathrm{\Delta V_{\alpha p}/V_A}$, and anisotropic $\mathrm{T_{\perp p}/T_{\parallel p}}$ that is mirror unstable. This part of PCS solar wind should come from closed loops deep inside the streamer belt probably via successive magnetic reconnection processes, which leads to the preferential heating of protons in perpendicular directions and then drives the mirror instability. In comparison, the other population has low but higher $\mathrm{A_{He}}$, much higher $\mathrm{\Delta V_{\alpha p}/V_A}$, and isotropic $\mathrm{T_{\perp p}/T_{\parallel p}}$. It means this population of PCS solar wind could originate from closed regions of streamer belt through magnetic reconnections, but the loops could be higher and less reconnection processes may be needed to release the plasma, which may create some waves or turbulence that are favorable to further accelerate the alpha particles, but the exact reasons need a detailed investigation.

\section{E4-E10 Results \label{sec:Eallresults}}

\begin{figure}
\epsscale{1.2}
\plotone{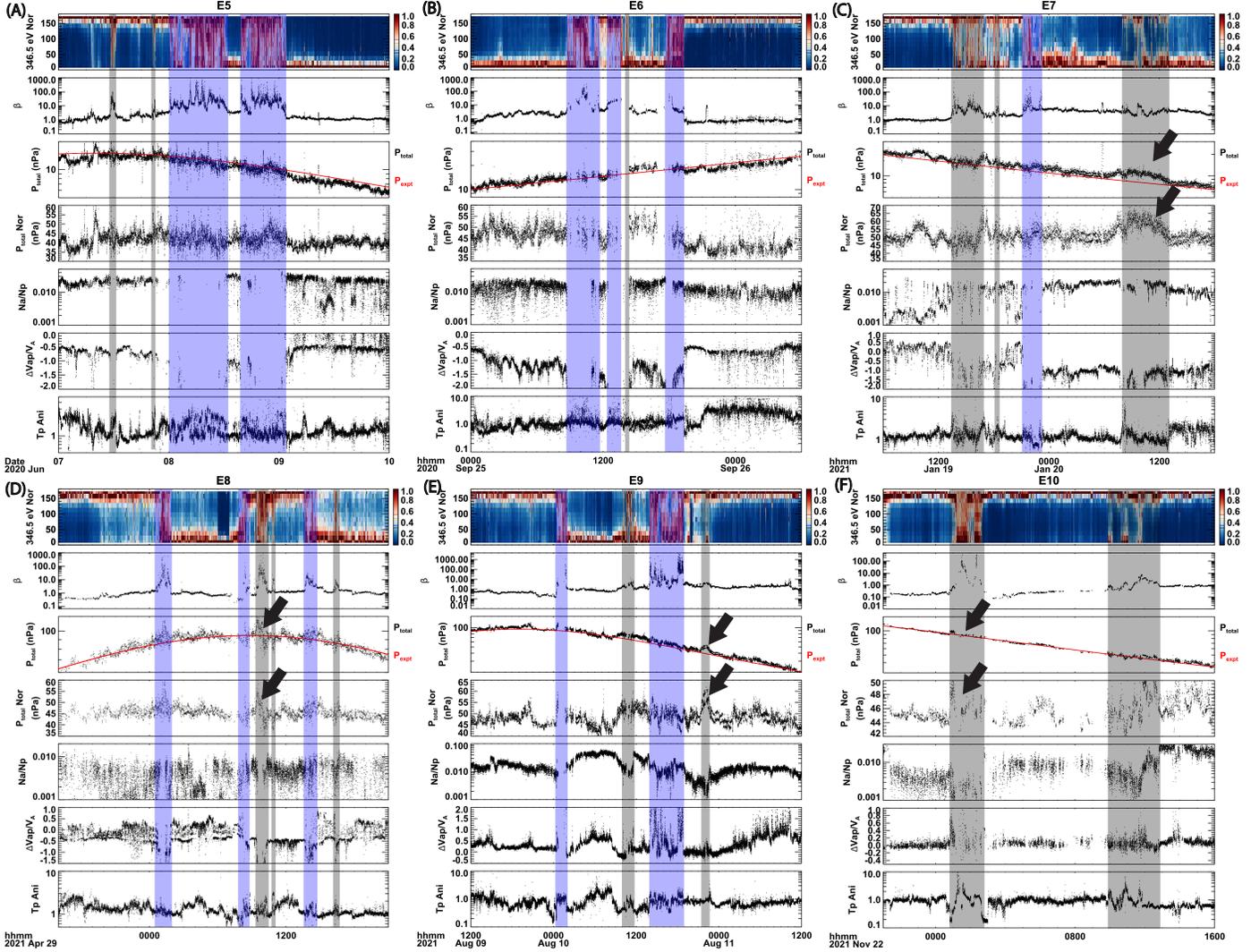}
\caption{Overview of high beta streamer belt solar wind from E5 to E10. In each figure, the panels from top to bottom show the normalized e-PAD at energy of 346.5 eV, plasma beta $\mathrm{\beta}$, total pressure and expected pressure, the normalized total pressure, alpha to proton abundance ratio $\mathrm{Na/Np}$, alpha-proton differential speed normalized by local Alfvén speed $\mathrm{\Delta Vap/V_A}$, and proton temperature anisotropy. The blue shaded regions mark the HPS solar wind, whereas the gray shaded regions mark the PCS solar wind. 
} \label{fig:Eallover}
\end{figure}
%%%%%%%%%%%%%%%%%%

In this section, we extend above study to include E5 to E10 observations. This is valuable to figure out whether the differences between HPS solar wind and PCS solar wind is of significance.

Figure \ref{fig:Eallover} gives an overview of the high $\beta$ streamer belt solar wind from E5 to E10. In each figure, the panels from top to bottom display the normalized e-PAD, $\mathrm{\beta}$, total pressure and expected pressure, the normalized total pressure, $\mathrm{Na/Np}$, $\mathrm{\Delta Vap/V_A}$, and proton temperature anisotropy. The HPS solar wind and PCS solar wind are marked by blue and gray shaded regions, respectively. The details of the high beta streamer belt solar wind intervals are listed in Table \ref{tab:HBSWs} in Appendix \ref{sec:SBSWlist}.

\subsection{Pressure variations \label{sec:EallPressure}}

\begin{figure}
\epsscale{1.1}
\plotone{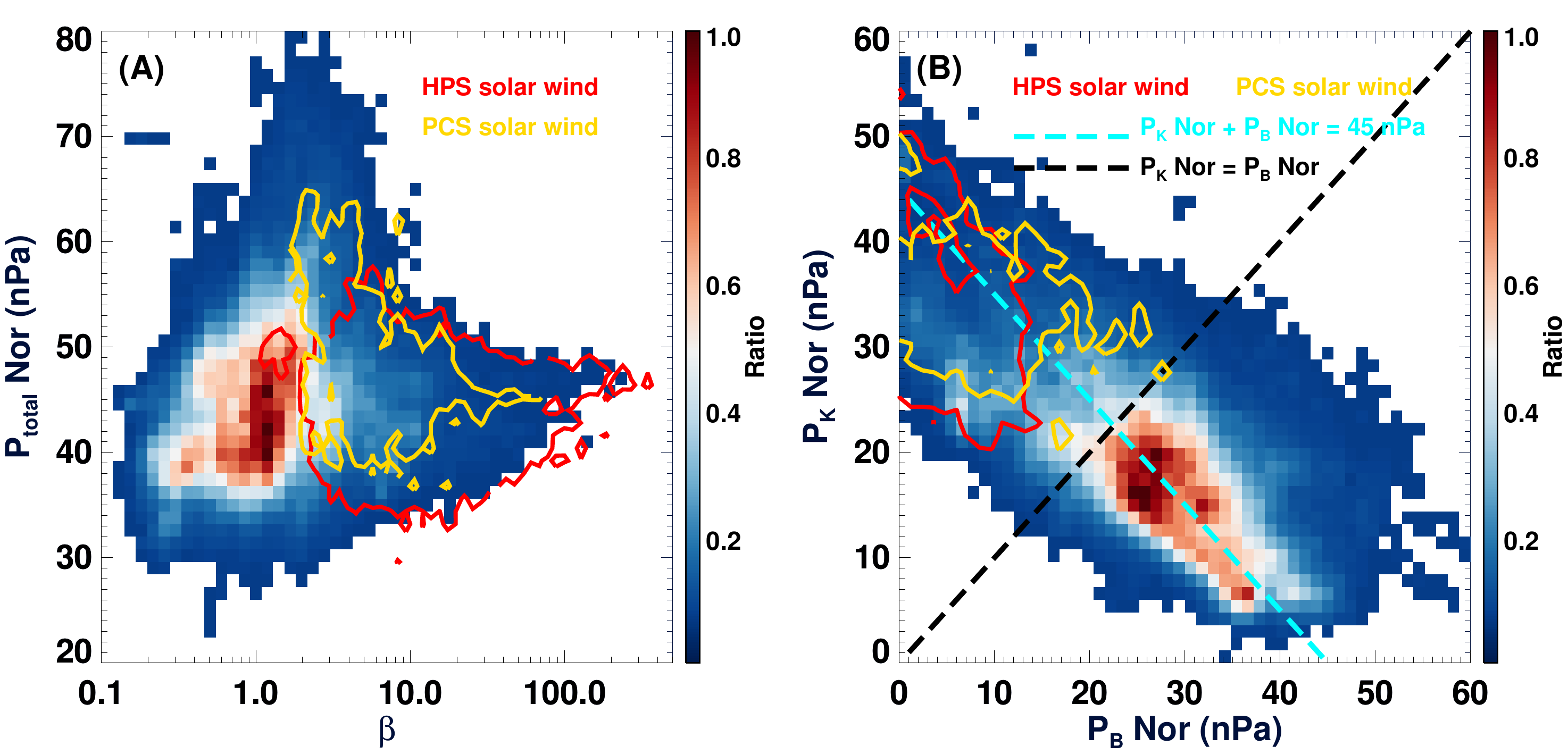}
\caption{Pressure variations in E4-E10. Panel (A) shows the distribution of normalized total pressure ($\mathrm{P_{total}\ Nor}$) versus plasma beta ($\mathrm{\beta}$). Panel (B) shows the distribution of normalized thermal pressure ($\mathrm{P_{k}\ Nor}$) versus normalized magnetic pressure ($\mathrm{P_{B}\ Nor}$). In both panels, the colors mean the occurrence ratios of solar wind. The red and gold contours indicate the HPS and PCS solar winds, respectively. In panel (B), the cyan dashed line represents $\mathrm{P_{B}\ Nor + P_{k}\ Nor = 45\ nPa}$, and the black dashed line suggest $\mathrm{P_{B}\ Nor = P_{k}\ Nor}$.
} \label{fig:EallPtotal}
\end{figure}
%%%%%%%%%%%%%%%%%%

In this part, we will show that the non-pressure-balanced signature of PCS solar wind is evidential. 

In Figure \ref{fig:Eallover}, we mark several PCS solar wind intervals from E7 to E10 with black arrows. These intervals display profound pressure enhancements, which can be seen both from the comparison between $\mathrm{P_{total}}$ and $\mathrm{P_{expt}}$ in the third panel and from the $\mathrm{P_{total}\ Nor}$ in the fourth panel in each figure. These distinct enhancements suggest that the PCS solar wind could be non-pressure-balanced structures.

Figure \ref{fig:EallPtotal} presents more details on the pressure variations in HPS solar wind and PCS solar wind. Panel (A) shows the distribution of $\mathrm{\beta}$ versus $\mathrm{P_{total}\ Nor}$. Panel (B) exhibits the distribution of $\mathrm{P_{k}\ Nor}$ versus $\mathrm{P_{B}\ Nor}$. In both panels, the colors indicate the occurrence ratios of solar wind in E4-E10 below 0.25 AU. The red and gold contours represent the HPS solar wind and PCS solar wind as listed in Table \ref{tab:HBSWs}, respectively. In panel (B), the cyan dashed line represents $\mathrm{P_{B}\ Nor + P_{k}\ Nor = 45\ nPa}$, and the black dashed line suggests $\mathrm{P_{B}\ Nor = P_{k}\ Nor}$.

From Figure \ref{fig:EallPtotal}(A), we can see the HPS solar wind generally has large $\mathrm{\beta}$ from about one to several hundreds, but $\mathrm{P_{total}\ Nor}$ is around 45 nT with the distributions spread below $\mathrm{\beta \sim 10}$. In PCS solar wind, $\mathrm{\beta}$ is smaller, which varies from 1 to about 60. $\mathrm{P_{total}\ Nor}$ is around 45 nT above $\mathrm{\beta \sim 6}$, but it is much larger than 45 nT below $\mathrm{\beta \sim 6}$, which further support above statement that the PCS solar wind should be non-pressure-balanced structure. However, the origin of the pressure enhancements from $\mathrm{\beta \sim 1}$ to $\mathrm{\beta \sim 5}$ in the background solar wind is unknown, which may be the unidentified PCS solar wind because the criteria to select PCS solar wind is generally strict in this work.
In Figure \ref{fig:EallPtotal}(B), we can see that $\mathrm{P_{K}\ Nor}$ is larger than $\mathrm{P_{B}\ Nor}$ in both HPS solar wind and PCS solar wind, as shown by the black dashed line, which is expected because magnetic filed strength usually depletes significantly around current sheet crossings. However, the total pressure enhancement in PCS solar wind should be caused by the less reduced or unreduced magnetic pressure therein as compared with HPS solar wind, as indicated by the cyan line. 

As a conclusion, the HPS solar wind is roughly pressure-balanced, but the PCS solar wind shows evidential non-pressure-balanced signature, and the enhancement of total pressure in PCS solar wind could be a result of the magnetic pressure does not reduce significantly.

\subsection{Temperature anisotropy characteristics \label{sec:EallTani}}

\begin{figure}
\epsscale{1.0}
\plotone{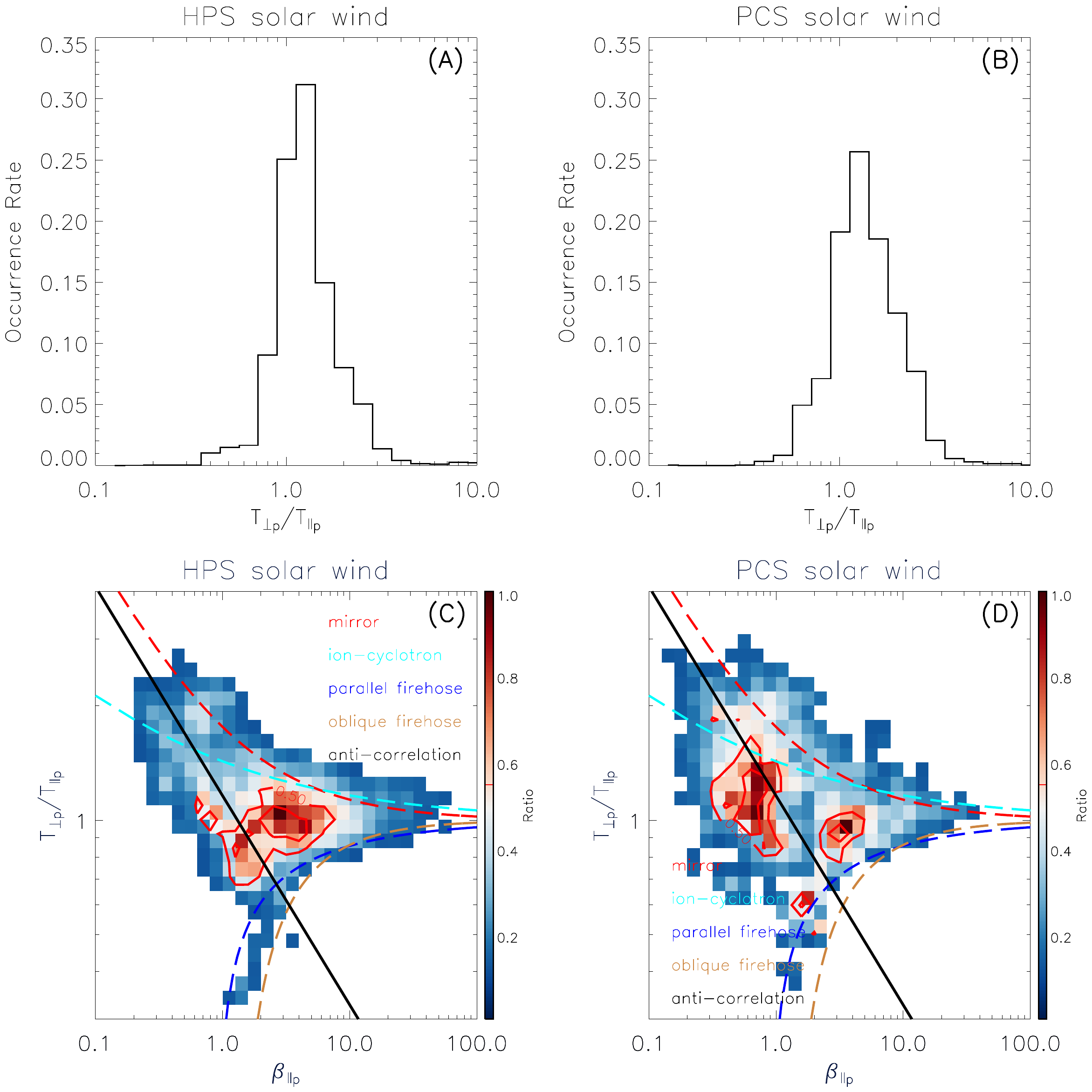}
\caption{Temperature anisotropies in streamer belt solar wind from E4 to E10. This figure has the same format as that in Figure \ref{fig:Tani}. 
} \label{fig:EallTani}
\end{figure}
%%%%%%%%%%%%%%%%%%

Figure \ref{fig:EallTani} shows the temperature anisotropies in streamer belt solar wind in E4-E10, with the same format as Figure \ref{fig:Tani}. This figure displays similar temperature anisotropy characteristics as indicated in Figure \ref{fig:Tani}. 

In HPS solar wind, we can see the $\mathrm{T_{\perp p}/T_{\parallel p}}$ is almost isotropic from panel (A), and the solar wind is well limited by the instabilities, with some exceeds the mirror limitation, as shown in panel (C). In comparison with E4 result, we can see the main distribution of $\mathrm{\beta_{\parallel p}}$ is larger, implying the $\mathrm{T_{\parallel p}}$ is higher in the following encounters when closer to the Sun, which is reasonable \citep{huang-2020}. These results are consistent with E4 result that the HPS solar wind is mostly in thermal equilibrium state.

In PCS solar wind, we can still see two populations in panel (D), but this signature is not noticeable in panel (B), which shows a broad distribution around $\mathrm{T_{\perp p}/T_{\parallel p} = 1}$. Additionally, the plasma is also limited by the instabilities, but some of the anisotropic population distribute beyond the mirror instability. This is also consistent with E4 result that some PCS solar wind is experiencing preferential proton heating.

\subsection{Helium signatures \label{sec:Eallhelium}}

\begin{figure}
\epsscale{1.1}
\plotone{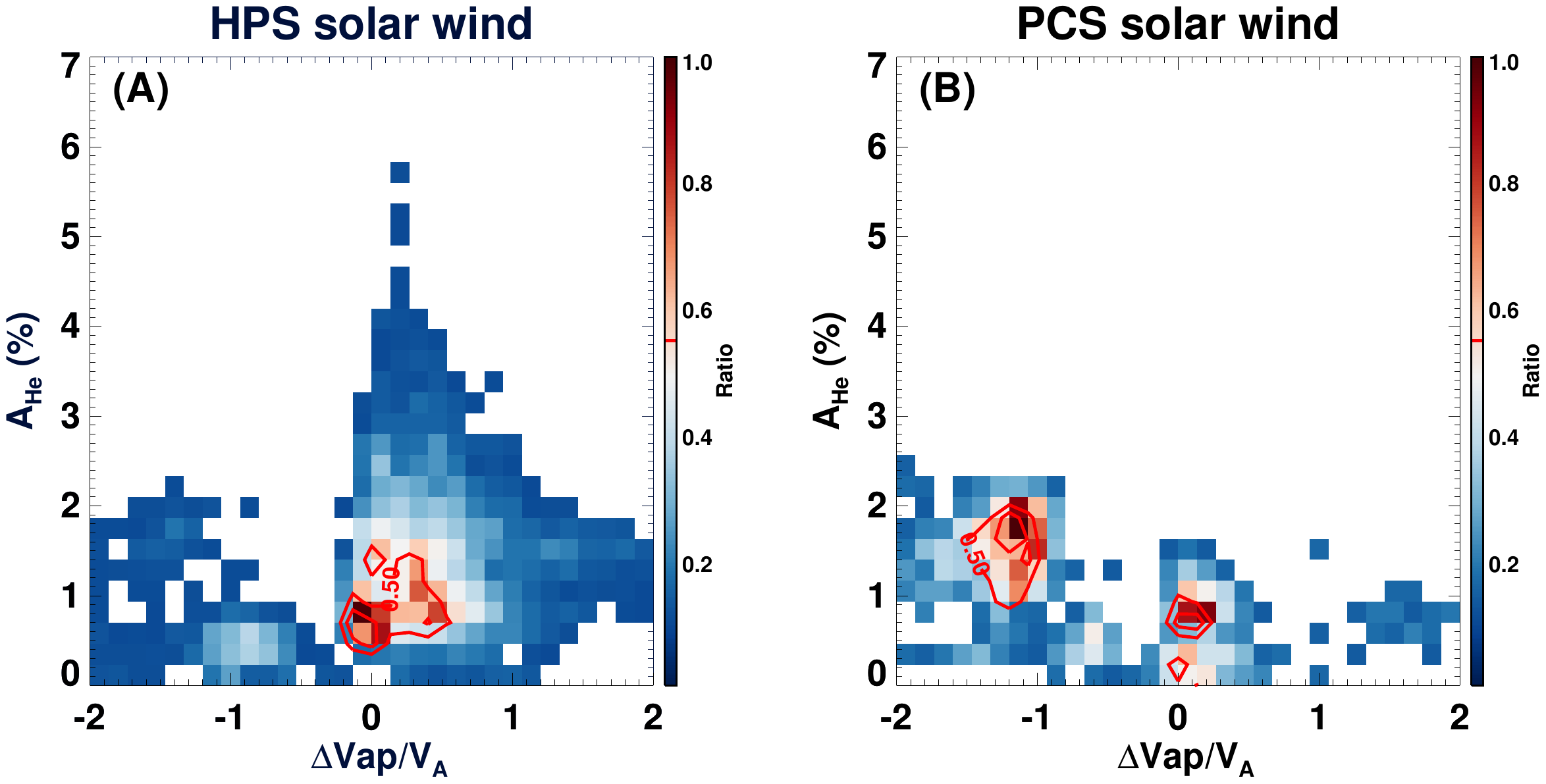}
\caption{The distributions of helium abundance ($\mathrm{A_{He}}$) versus alpha-proton differential speed that normalized by local Alfvén speed ($\mathrm{\Delta V_{\alpha p}/V_A}$) in streamer belt solar wind from E4 to E10. This figure has the same format as that in Figure \ref{fig:NVap}.  
} \label{fig:Eallhelium}
\end{figure}

\begin{figure}
\epsscale{1.1}
\plotone{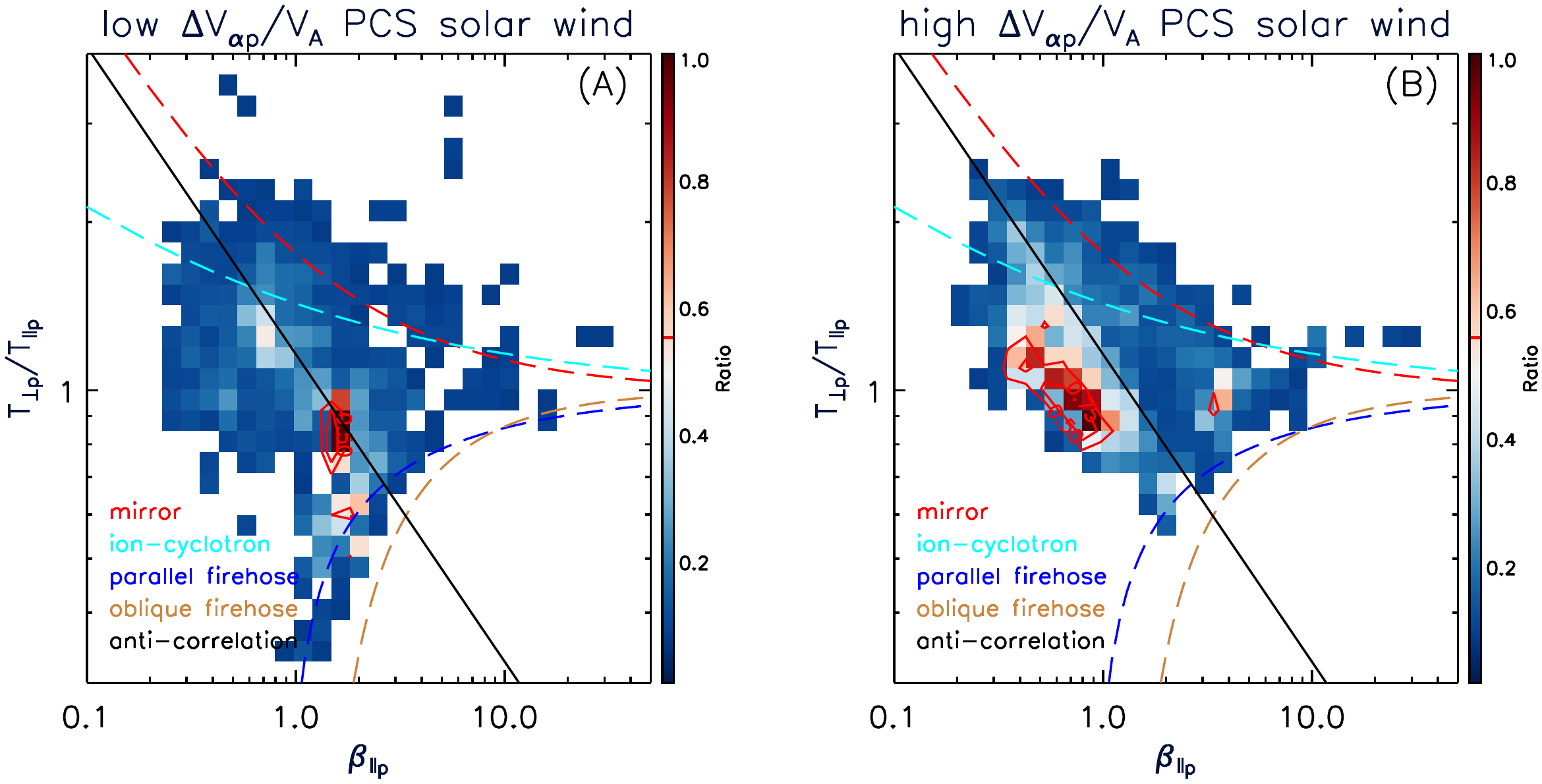}
\caption{Temperature anisotropy distributions of PCS solar wind with different $\mathrm{\Delta V_{\alpha p}/V_A}$ from E4 to E10. This figure has similar format as that in the bottom of Figure \ref{fig:Tani}. 
} \label{fig:EallPCSTani}
\end{figure}
%%%%%%%%%%%%%%%%%%

Figure \ref{fig:Eallhelium} displays the helium signatures in streamer belt solar wind in E4-E10, with the same format as Figure \ref{fig:NVap}. It also shows similar results as E4, but the distributions are more complicated.

In HPS solar wind, we can see the major distribution is dominated by low $\mathrm{A_{He}}$ ($\sim 1\%$) and low $\mathrm{\Delta V_{\alpha p}/V_A}$ (around 0). But we can also see low $\mathrm{A_{He}}$ and high $\mathrm{\Delta V_{\alpha p}/V_A}$ ($|\mathrm{\Delta V_{\alpha p}/V_A}| > 1$) population, and high $\mathrm{A_{He}}$ ($> 4\%$) and low $\mathrm{\Delta V_{\alpha p}/V_A}$ population. This indicates the HPS solar wind originates from both closed and open magnetic field regions at the Sun. The large positive $\mathrm{\Delta V_{\alpha p}/V_A}$ implies the alphas could be further accelerated \citep{isenberg-1983, kasper-2017}, but the negative values infer that the alphas may be decelerated over distance during solar wind expansion \citep{neugebauer-1994, maneva-2015, mostafavi-2022}. The complicated variations of helium signatures in HPS solar wind suggest that the selected HPS solar wind intervals may contain plasma from extra source regions, and this also indicates the complex variations of solar wind in the inner heliosphere.

In PCS solar wind, the two populations are still distinctive, but the low $\mathrm{A_{He}}$ and high $\mathrm{\Delta V_{\alpha p}/V_A}$ population shifts predominantly from positive $\mathrm{\Delta V_{\alpha p}/V_A}$ values to negative ones. Therefore, the PCS solar wind should mainly come from the closed magnetic field regions of the streamer belt, but the alphas could experience either acceleration or deceleration processes during propagation as stated above.
In addition, we note the linear relationship between $\mathrm{\Delta V_{\alpha p}/V_A}$ and $\mathrm{\beta_{\parallel p}}$ is roughly maintained (not shown) in PCS solar wind, but temperature anisotropy populations are somewhat overlapped with each other as shown in Figure \ref{fig:EallTani}. Thus, we display the temperature anisotropy distributions of PCS solar wind with different $\mathrm{\Delta V_{\alpha p}/V_A}$ in Figure \ref{fig:EallPCSTani}. Panel (A) shows PCS solar wind with low $\mathrm{\Delta V_{\alpha p}/V_A}$ ($\mathrm{|\Delta V_{\alpha p}/V_A}|<0.8$), whereas panel (B) presents another population with high $\mathrm{\Delta V_{\alpha p}/V_A}$ ($\mathrm{|\Delta V_{\alpha p}/V_A}|>0.8$). From this figure, we can see the low $\mathrm{\Delta V_{\alpha p}/V_A}$ population has broader $\mathrm{T_{\perp p}/T_{\parallel p}}$ distributions and slightly more anisotropic population beyond mirror instability. One more interesting feature regarding the high $\mathrm{\Delta V_{\alpha p}/V_A}$ PCS solar wind is that the major temperature anisotropies seem to distribute along the anti-correlation line, as shown by the black line in panel (B), which corresponds mainly to the proton core behaviors. This implies the low $\mathrm{\Delta V_{\alpha p}/V_A}$ PCS solar wind may involve with proton beam variations, and it is valuable to further investigate the differences of velocity distribution functions in the two populations of PCS solar wind. 

As a result, the multi-event study in this section suggests that the PCS solar wind is different from HPS solar wind. The features of HPS solar wind are generally consistent with that observed at 1 AU, but the temperature anisotropy and helium signatures are more complicated in the inner heliosphere. However, PCS solar wind shows non-pressure-balanced signature and has two populations that involve with preferential proton heating and/or alpha acceleration/deceleration.

\section{Discussion} \label{sec:disc}

Here, we want to clarify that the PCS solar wind should not be magnetic holes. We note that \citet{chenc-2021} use similar criteria, including reduced magnetic field, enhanced density, high plasma beta, and partial current sheet crossing, to identify this kind of solar wind as macro magnetic holes, which could be caused by the HCS ripples. The magnetic holes have been studied for several decades since they are first reported by \citet{turner-1977}. In general, the magnetic coronal holes are isolated, pressure-balanced structures with the magnetic field strength significantly reduced, and they are more often observed in fast solar wind or in environment with high $\mathrm{\beta}$ and $\mathrm{T_{\perp p}/T_{\parallel p}}$ \citep[e.g.,][and references therein]{stevens-2007, chenc-2021}. 

However, our result suggests that PCS solar wind should be non-pressure-balanced structures, which is contrast to the most distinguishing characteristic of magnetic holes. Moreover, in PCS solar wind, the kinetic pressure is larger and the magnetic pressure is smaller than that in ambient solar wind. But the total pressure enhancement in PCS solar wind is mainly caused by the less decreased magnetic pressure as compared with HPS solar wind, as shown in Figure \ref{fig:EallPtotal}(B), this further indicates that the magnetic field in PCS solar wind does not reduce that significantly, which is also contrast to the primary definition of magnetic holes. In addition, as discussed by \citet{phan-2021}, the partial current sheet crossings are likely generated by the traveling large plasma blobs bulging onto both sides of the HCS crossings based on their signatures of long duration and recurrent appearance. Thus, the partial current sheet crossings may not associate with rippled HCS. As a result, we suggest the PCS solar wind should not be identified as magnetic holes especially its source is clear, this is important to correctly understand the solar wind properties. Additionally, the investigations of kinetic-scale and small-scale magnetic holes \citep[e.g.,][]{huangsy-2021, yu-2021holes, yu-2022holes, zhou-2022} may need to verify whether they are associated with PCS solar wind. 

Moreover, we discuss some of the slow solar wind observations by PSP that associated with this work. We note that the four years of discoveries at solar cycle minimum by PSP has been thoroughly reviewed by \citet{raouafi-2023}. As the PSP approaches to the solar atmosphere, new features of slow solar wind have been uncovered near the Sun. First, PSP has detected several sub-Alfvénic solar wind intervals since it entered the Alfvén critical surface in E8 \citep{kasper-2021, bandy-2021, zank-2022, zhao-2022}. However, the selected streamer belt solar wind intervals have no overlap with the sub-Alfvénic solar wind intervals, due to streamer belt solar wind is featured by high plasma beta whereas sub-Alfvénic solar wind is characterized by low plasma beta. Further, based on our preliminary analysis of four long intervals of sub-Alfvénic solar wind \footnote{Interval 1 (E8): 2021-04-28 09:33 -- 2021-04-28 14:42 UT; Interval 2 (E9): 2021-08-09 21:30 -- 2021-08-10 00:24 UT; Interval 3 (E10): 2021-11-21 21:23 -- 2021-11-22 00:57 UT; Interval 4 (E10): 2021-11-22 02:40 -- 2021-11-22 09:52 UT.}, we suggest the sub-Alfvénic solar wind could be pressure-balanced structures, implying they are well evolved solar wind streams that probably originate from pseudostreamers in the Sun \citep{kasper-2021}. Second, PSP observes prevalent Alfvénic slow solar wind in the past four years. Alfvénic slow solar wind is slow wind with high Alfvénicity, which is a typical feature for fast solar wind rather than the slow one. Alfvénic slow solar wind is not common to be observed at 1 AU, and studies suggest that they should come from the coronal holes in the Sun \citep{dAmicis-2015, dAmicis-2018, wangx-2019}. PSP observations indicate pervasive Alfvénic slow solar wind in the inner heliosphere, and case studies confirm their origins from coronal holes \citep{bale-2019, griton-2021}. However, \citet{huang-2020ASSW} shows that highly Alfvénic slow solar wind shares similar temperature anisotropy and helium abundance properties with regular slow solar wind, and they thus may have multiple origins based on statistical analysis. In addition, current sub-Alfvénic solar wind observed by PSP is also Alfvénic slow solar wind \citep{zank-2022, zhao-2022}. The streamer belt solar wind generally has low Alfvénicity \citep{huang-2020ASSW, zhao-2022}, and it would be valuable to further disclose the differences of slow solar wind with different Alfvénicities. Third, the slow solar wind is very dynamic in the near-Sun environment as introduced in section \ref{sec:intro}. The spatial and temporal variabilities of slow solar wind are further increased due to multiple current sheet crossings \citep{szabo-2019, lavraud-2020}, magnetic reconnection exhausts \citep{phan-2021}, small flux ropes \citep{chenyu-2021, zhao-2021, reville-2022, chenyu-2023}, turbulences \citep{chen-2021, zank-2022, zhao-2022}, switchbacks \citep{kasper-2019, de-2020, fisk-2020, horbury-2020, zank-2020}, and so on. As a result, investigating the slow solar wind that either originate from the same source region or share similar properties could reduce the uncertainty in such studies, and the pressure, temperature anisotropy and alpha characteristics could be crucial to understand the underlying mechanisms of slow solar wind. 
%The origin of slow solar wind from the Sun is unclear, which has long been one of the major unsolved problems in heliospheric physics \citep{antiochos-2011}, due to two major difficulties: highly spatial and temporal variabilities, and larger angular extent than expected from streamer belt solar wind \citep{wang-2000, abbo-2016}. Investigating the origin of solar wind is thus an important science objective of PSP mission. 

\section{Summary} \label{sec:sum}
In this work, using the PSP observations from E4 to E10, we identify streamer belt solar wind from enhancements in plasma beta, and we further use electron pitch angle distributions to separate it into HPS solar wind that around the full HCS crossings and PCS solar wind that in the vicinity of PCS crossings. Focusing on E4 observations, we find the two kinds of solar wind show different characteristics of pressure, temperature anisotropy and helium distributions. By extending this study to E10, we figure out more complicated variations of above parameters therein. The major results are summarized in the following.

\begin{enumerate}
\item The HPS solar wind is generally pressure-balanced, but the PCS solar wind should be non-pressured-balanced structures. The total pressure of PCS solar wind increases evidently, which is caused by the fact that the magnetic pressure therein does not reduce significantly as compared with HPS solar wind.

\item The HPS solar wind is mostly in thermal equilibrium state, but PCS solar wind has two populations. One population of PCS solar wind has isotropic proton temperatures, but the other population shows anisotropic signature with some solar wind being mirror unstable. 

\item The HPS solar wind is characterized by low $\mathrm{\Delta V_{\alpha p}/V_A}$, whereas the PCS solar wind is dominated by low $\mathrm{A_{He}}$. The HPS solar wind shows low $\mathrm{\Delta V_{\alpha p}/V_A}$ but its $\mathrm{A_{He}}$ covers low to high values. However, the PCS solar wind has two populations, with one population distinguished by low $\mathrm{A_{He}}$ and low $\mathrm{\Delta V_{\alpha p}/V_A}$ whereas the other population displaying low $\mathrm{A_{He}}$ but high $\mathrm{\Delta V_{\alpha p}/V_A}$. Further, the low $\mathrm{A_{He}}$ and low $\mathrm{\Delta V_{\alpha p}/V_A}$ population relates to anisotropic temperatures, but the low $\mathrm{A_{He}}$ and high $\mathrm{\Delta V_{\alpha p}/V_A}$ population is almost isotropic. Furthermore, the multi-event study reveals more complicated variations in inner heliosphere.

\end{enumerate}

Combining all of the observations, we can conclude that the HPS solar wind is similar to that observed at 1 AU and beyond, which is pressure-balanced structure with thermal equilibrium state and regular helium signature, implying the HPS solar wind comes from both closed and open magnetic field regions of streamer belt and it is generally well evolved. However, the PCS solar wind is non-pressure-balanced structure that has two populations. One population exhibits very low $\mathrm{A_{He}}$, low $\mathrm{\Delta V_{\alpha p}/V_A}$, and anisotropic $\mathrm{T_{\perp p}/T_{\parallel p}}$ that is mirror unstable, implying it originates from closed loops deep inside the streamer belt probably via successive magnetic reconnection processes, which preferentially heats protons in perpendicular directions and then possibly drives the mirror instability. In comparison, the other population has low but higher $\mathrm{A_{He}}$, much higher $\mathrm{\Delta V_{\alpha p}/V_A}$, and isotropic $\mathrm{T_{\perp p}/T_{\parallel p}}$, suggesting this population is from closed regions of streamer belt through magnetic reconnections, but the loops locate at higher altitude and less reconnections are needed to release the plasma. Consequently, we draw another conclusion that the PCS solar wind should not be magnetic holes.

%%%%%%%%%%%%%%%%%%%%%%%%%%%%%%%%%%%%%%%%%%%%
\appendix
\section{Parameters} \label{sec:parameters}
The electron pressure $\mathrm{P_{e} = n_e k_B T_e}$, proton and alpha pressure $\mathrm{P_{p+\alpha} = n_p k_B T_p + n_{\alpha} k_B T_{\alpha}}$, total kinetic pressure $\mathrm{P_{k} = P_{e} + P_{p+\alpha}}$, magnetic pressure $\mathrm{P_{B} = B^2/2\mu_0}$, and total pressure $\mathrm{P_{total} = P_{k} + P_{B}}$. In the equations, $\mathrm{\mu_0}$ is vacuum magnetic permeability, $\mathrm{k_B}$ is Boltzmann constant, $B$ is the magnetic field strength, $n_i$ and $T_i$ are the number density and temperature of particle \textit{i} species, where \textit{i} is e, p, and $\alpha$ for electron, proton and alpha particle, respectively. 

The subscripts ${\perp}$ and ${\parallel}$ represent the perpendicular and parallel directions with respect to ambient magnetic field \textbf{B}. $T_{\perp p}$, $T_{\parallel p}$, and $T_{\perp p}/T_{\parallel p}$ are the perpendicular proton temperature, parallel proton temperature, and proton temperature anisotropy, respectively. $\beta_{\parallel p} = 2\mu_0 n_p k_B T_{\parallel p}/B^2$ is the parallel proton beta, whereas $\mathrm{\beta} = 2\mu_0 n_p k_B T_p/B^2$ is the plasma beta. 

In addition, $B_r$ and $R_S$ are radial component of magnetic field and heliocentric distance, respectively. $N_\alpha/N_p$ is the alpha to proton number density ratio, and $A_{He} = N_\alpha/N_p \times 100\%$ measures the helium abundance ratio. Moreover, the alpha-proton differential speed is $\mathrm{\Delta V_{\alpha p}} = (v_{\alpha r}-v_{pr})/cos(\theta)$, where $v_{\alpha r}$ and $v_{pr}$ are the radial speeds of alpha particle and proton, respectively, and $cos(\theta) = |B_r/B|$ is used to assure the derived differential speed is independent with magnetic field polarity \citep{reisenfeld-2001, fu-2018}. 
Besides, the local Alfvén speed is calculated with $\mathrm{V_A}=|B|/\sqrt{\mu_0(N_pm_p+N_{\alpha}m_{\alpha})}$, where $m_p$ and $m_{\alpha}$ are the mass of proton and alpha particle, respectively. In the calculations, we use the electron density derived from the analysis of plasma quasi-thermal noise (QTN) spectrum measured by the FIELDS Radio Frequency Spectrometer \citep{pulupa-2017, moncuquet-2020} to replace $N_p$ by assuming $\mathrm{A_{He}}$ is 4\%, which does not significantly change the $\mathrm{V_A}$ as $\mathrm{A_{He}}$ generally varies from 1\% to 8\% \citep{liu-2021, mostafavi-2022}.

\section{Radial Evolution of Pressures} \label{sec:presevo}

\begin{figure}
\epsscale{0.70}
\plotone{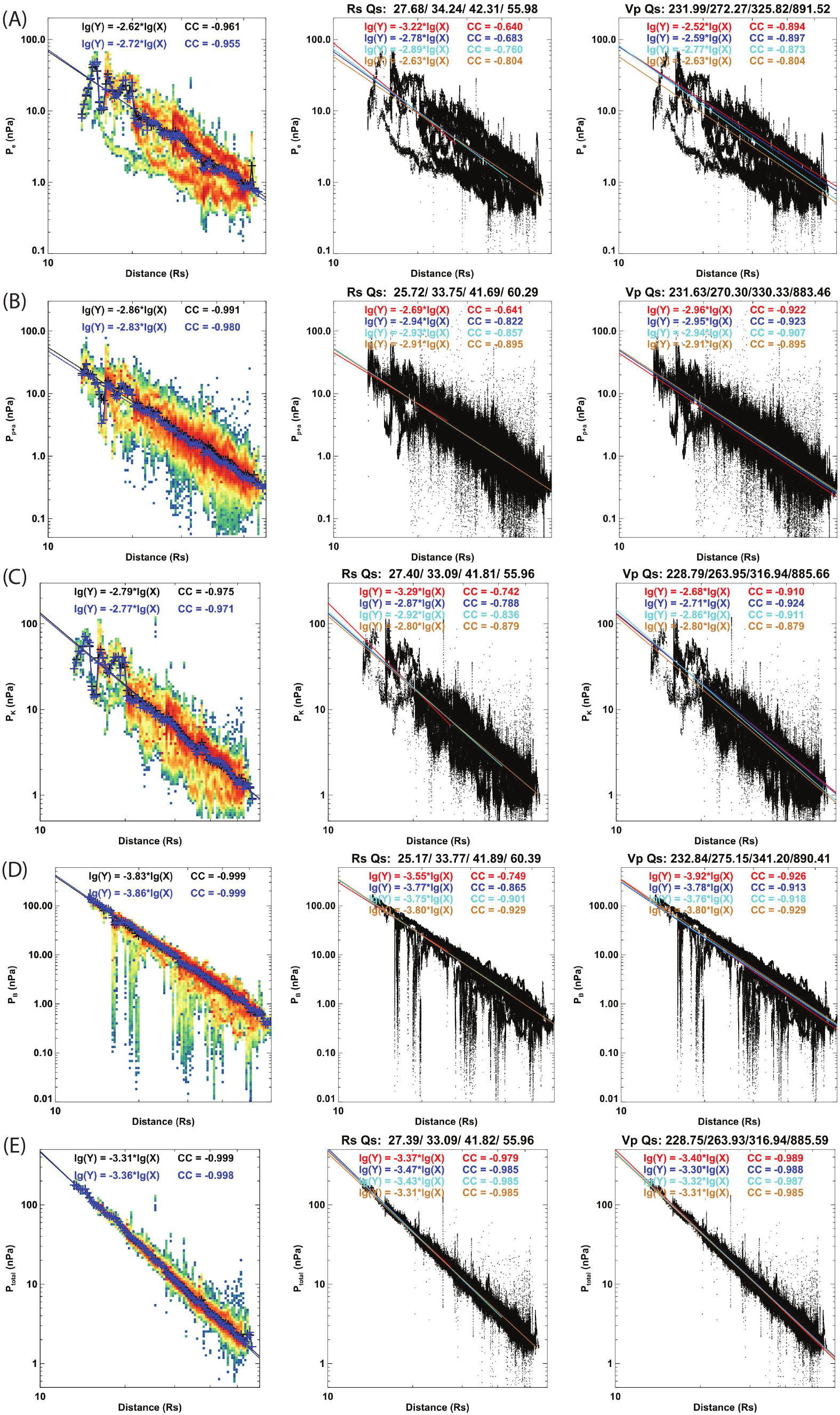}
\caption{The radial evolutions of pressure components of all solar wind from E4 to E12 below 0.25 AU. From top to bottom, rows (A) to (E) present the radial evolution of electron pressure $\mathrm{P_{e}}$, proton and alpha pressure $\mathrm{P_{p+\alpha}}$, total kinetic pressure $\mathrm{P_{k}}$, magnetic pressure $\mathrm{P_{B}}$, and total pressure $\mathrm{P_{total}}$, respectively.
In each row, the left panel shows the pressure component with the color indicating the occurrence ratio of data points. The black and blue lines represent the fitted results based on the mean and median values of the pressure component at each distance bin, respectively. The fitted evolution indices and the correlate coefficients are presented accordingly. 
The middle and right panels show the fittings based on the four distance quantiles and four speed quantiles, respectively. In both panels, the red, blue, cyan and brown lines indicate the fitting results according to the first (25\%), second (50\%), third (75\%) and fourth (100\%) quantile, with the fitted evolution indices and the correlate coefficients presenting accordingly.  
\label{fig:PPall}}
\end{figure}
%%%%%%%%%%%%%%%%%%%%%%%%%%%%%%%%%

Figure \ref{fig:PPall} shows the radial evolutions of pressure components of all solar wind from E4 to E12 below 0.25 AU. From top to bottom, rows (A) to (E) present the radial evolution of electron pressure $\mathrm{P_{e}}$, proton and alpha pressure $\mathrm{P_{p+\alpha}}$, total kinetic pressure $\mathrm{P_{k}}$, magnetic pressure $\mathrm{P_{B}}$, and total pressure $\mathrm{P_{total}}$, respectively. 
In each row, the left panel shows the pressure component with the color indicating the occurrence ratio of data points. The black and blue lines represent the fitted results based on the mean and median values of the pressure component at each distance bin, respectively. The fitted evolution indices and the correlate coefficients are presented accordingly. 
The middle and right panels show the fittings based on the four distance quantiles (as shown in the top with unit being of $\mathrm{R_S}$) and four speed quantiles (as shown in the top with unit being of $\mathrm{km/s}$), respectively. In both panels, the red, blue, cyan and brown lines indicate the fitting results according to the first (25\%), second (50\%), third (75\%) and fourth (100\%) quantile, with the fitted evolution indices and the correlate coefficients presenting accordingly. We use a power law function to fit the radial evolution of each pressure component. 

From this figure, we can see the radial evolution of pressure components varies with both heliocentric distance and solar wind speed, but the fitting results are comparable and the correlation coefficients are pretty high. In this work, we don't focus on their variations with different distance range and different speed range, thus we select the fitting results with higher correlation coefficients in left panel in each row for normalization. Consequently, we choose the power law indices of $\mathrm{P_{e}}$, $\mathrm{P_{p+\alpha}}$, $\mathrm{P_{k}}$, $\mathrm{P_{B}}$, and $\mathrm{P_{total}}$ to be -2.62, -2.86, -2.79, -3.83, and -3.31, respectively. 

Moreover, the ideal spherical adiabatic expansion predicts that the magnetic field strength $B$ and ion density $n$ decrease as $R^{-2}$ for a solar wind expanding with constant speed. Further, with an ideal polytropic index $\gamma = 5/3$, the kinetic pressure and total temperature follows the relationship of $\mathrm{P_{k}} \propto n^{\gamma} \propto R^{-10/3}$ and $T \propto n^{\gamma-1} \propto R^{-4/3}$, respectively, whereas the ion pressures approximately follow $\mathrm{P_{i}} \propto R^{-10/3}$ as $\mathrm{P_k}$ and the magnetic pressure follows $\mathrm{P_{B}} \propto B^2 \propto R^{-4}$. For the total pressure $\mathrm{P_{total} = P_k + P_B}$, we can derive that its radial power law index should vary between that for the $\mathrm{P_{B}}$ and $\mathrm{P_{k}}$, i.e. -4 to -10/3. As the solar wind in the interplanetary space generally has plasma beta $\beta > 1$, the radial power law index of $\mathrm{P_{total}}$ is expected to be close to -10/3.

In comparison with our fitting results, we can see $\mathrm{P_{total}} \propto R^{-3.31}$ and $\mathrm{P_{B}} \propto R^{-3.83}$ are very close to the prediction of adiabatic expansions, but the ion pressure components and total kinetic pressure show flatter slopes. This indicates the solar wind observed by PSP is generally experiencing adiabatic expansion in the inner heliosphere, however the ion pressures do not match well with the adiabatic expansion predictions, which is possibly caused by the stronger anisotropic heating of ions in the near Sun environment \citep{huang-2020}.

\section{The streamer belt solar wind intervals} \label{sec:SBSWlist}
%%%%\startlongtable
\begin{deluxetable}{|c|c|c|c|c|}
\tablecaption{The high beta streamer belt solar wind intervals in each encounter. \label{tab:HBSWs}}
\tablecolumns{30}
\tablenum{1}
\tablewidth{750 pt}
\tablehead{
\multicolumn{1}{|c|}{Encounter} & \multicolumn{1}{|c|}{NO.} & \multicolumn{1}{|c|}{Start Time (UT)} & \multicolumn{1}{|c|}{End Time (UT)} & \multicolumn{1}{|c|}{Current Sheet Crossing} 
}
\startdata
E4	&  01 & 2020-01-30/13:40:00	&  2020-01-30/16:54:20	&  PCS  \\
	&  02 & 2020-01-31/19:56:00	&  2020-01-31/23:52:00	&  PCS  \\
	&  03 & 2020-02-01/02:44:00	&  2020-02-01/20:12:00	&  HCS  \\
\hline 
E5	&  04 & 2020-06-07/11:21:00	&  2020-06-07/12:24:00	&  PCS  \\
	&  05 & 2020-06-07/20:25:00	&  2020-06-07/21:09:00	&  PCS  \\
	&  06 & 2020-06-08/00:12:00	&  2020-06-08/12:40:00	&  HCS  \\
	&  07 & 2020-06-08/15:42:00	&  2020-06-09/01:32:00	&  HCS  \\
\hline  
E6	&  08 & 2020-09-25/08:42:00	&  2020-09-25/11:42:00	&  HCS  \\
	&  09 & 2020-09-25/12:26:20	&  2020-09-25/13:49:10	&  HCS  \\
	&  10 & 2020-09-25/13:52:20	&  2020-09-25/14:41:30	&  PCS  \\
	&  11 & 2020-09-25/17:40:30	&  2020-09-25/18:28:40	&  HCS  \\
\hline 
E7	&  12 & 2021-01-19/13:31:00	&  2021-01-19/16:46:00	&  PCS  \\
	&  13 & 2021-01-19/18:16:00	&  2021-01-19/18:31:00	&  PCS  \\
	&  14 & 2021-01-19/21:08:30	&  2021-01-19/23:26:00	&  HCS  \\
	&  15 & 2021-01-20/07:55:30	&  2021-01-20/13:32:00	&  PCS  \\
\hline 
E8	&  16 & 2021-04-29/00:44:50	&  2021-04-29/01:51:10	&  HCS  \\
	&  17 & 2021-04-29/08:14:40	&  2021-04-29/08:51:30	&  HCS  \\
	&  18 & 2021-04-29/09:24:40	&  2021-04-29/10:22:40	&  PCS  \\
	&  19 & 2021-04-29/10:48:10	&  2021-04-29/10:57:30	&  PCS  \\
	&  20 & 2021-04-29/13:40:10	&  2021-04-29/14:23:40	&  HCS  \\
	&  21 & 2021-04-29/16:15:40	&  2021-04-29/16:38:20	&  PCS  \\
\hline 
E9	&  22 & 2021-08-10/00:30:00	&  2021-08-10/01:54:00	&  HCS  \\
	&  23 & 2021-08-10/10:34:00	&  2021-08-10/11:38:00	&  PCS  \\
	&  24 & 2021-08-10/13:52:00	&  2021-08-10/19:50:00	&  HCS  \\
	&  25 & 2021-08-10/21:43:00 &  2021-08-10/22:56:00  &  PCS  \\
\hline 
E10	&  26 & 2021-11-22/01:10:00	&  2021-11-22/02:37:00	&  PCS  \\
	&  27 & 2021-11-22/09:58:00	&  2021-11-22/12:10:40	&  PCS  \\
\enddata
\end{deluxetable}

Table \ref{tab:HBSWs} lists all of the high beta streamer belt solar wind intervals from E4 to E10 as shown in Figure \ref{fig:overview} and Figure \ref{fig:Eallover}. From left to right, the columns indicate the PSP encounter, the number of selected interval, the start time, the end time, and the type of current sheet crossings. As a summary, there are 12 HCS crossings and 15 PCS crossings.

%%%%%%%%%%%%%%%%%%%%%%%%%%%%%%%%%%%%%%%%%%%%%%%%%%%%%%%%%%%%%%%%
\acknowledgments

Parker Solar Probe was designed, built, and is now operated by the Johns Hopkins Applied Physics Laboratory as part of NASA’s Living with a Star (LWS) program (contract NNN06AA01C). Support from the LWS management and technical team has played a critical role in the success of the Parker Solar Probe mission.
Thanks to the Solar Wind Electrons, Alphas, and Protons (SWEAP) team for providing data (PI: Justin Kasper, BWX Technologies). Thanks to the FIELDS team for providing data (PI: Stuart D. Bale, UC Berkeley). Jia Huang is also supported by NASA grant 80NSSC22K1017. Lan K. Jian is supported by LWS research program. 

\bibliography{HBSSW}{}
\bibliographystyle{aasjournal}

\end{document}